\documentclass[10pt,conference]{IEEEtran}
\IEEEoverridecommandlockouts

\usepackage[letterpaper, left=0.65in,right=0.65in, top=0.75in,bottom=1.1in]{geometry}

\usepackage{amssymb,amsfonts, amsmath, amsthm}
\usepackage{mathabx} 

\usepackage{xcolor}
\usepackage{epsfig}
\usepackage{stackengine}
\usepackage{color}
\usepackage{graphicx}
\usepackage{subcaption}
\usepackage{bm}
\usepackage{balance}
\usepackage{romannum}
\usepackage{textcomp}
\newcommand\norm[1]{\left\lVert#1\right\rVert}
\newtheorem{theorem}{Theorem} 
\newtheorem{lemma}{Lemma} 
\newtheorem{corollary}{Corollary} 
\newtheorem{proposition}{Proposition} 
\usepackage{algorithm,algpseudocode}
\algrenewcommand\algorithmicindent{0.7em}
\usepackage{cite}
\usepackage{bbm}
\usepackage{physics}
\def\calC{{\cal C}}
\def\calU{{\cal U}}

\def\PUN{{\cal PU}_n}
\def\SUN{{\cal SU}_n}
\def\UN{{\cal U}_n}
\def\U1{{\cal U}_1}
\def\matU{\mathbf{U}}
\def\matV{\mathbf{V}}
\def\matI{\mathbf{I}}

\def\be{\begin{equation}}
\def\ee{\end{equation}}

\newcommand{\diag}{{\mathrm{diag}}}

\def\BibTeX{{\rm B\kern-.05em{\sc i\kern-.025em b}\kern-.08em
    T\kern-.1667em\lower.7ex\hbox{E}\kern-.125emX}}

\begin{document}

\title{Distance and Distortion Bounds in the Projective Unitary Group with Respect to Chordal Distance}

\author{
       \IEEEauthorblockN{Bhanu Pratap Yadav, Mahdi~Bayanifar,  and  Olav~Tirkkonen }
             \\ \IEEEauthorblockA{\em Department of Information and Communications Engineering, Aalto University, Finland\\
              Email: \{bhanu.yadav, mahdi.bayanifar, olav.tirkkonen\}@aalto.fi, 
              }
}
\maketitle
\begin{abstract}
We consider the geometry of the projective unitary group $\PUN$ 
induced by a  global phase invariant  chordal distance, relevant for universal quantum computing. We obtain the volume and measure of small metric balls in $\PUN$ and derive the corresponding Gilbert-Varshamov
and Hamming bounds.  We provide upper and lower bounds for the kissing radius of codebooks in $\PUN$ as a function of the minimum distance. Using the lower bound of the kissing radius, we find
a tight Hamming bound. In addition, we establish bounds on the rate-distortion function for discretizing a source uniformly distributed
over $\PUN$ with a codebook of given cardinality, and provide the expected value of the covering radius
when the cardinality of the code is asymptotically large. For any code in $\PUN$ with given  cardinality we provide a lower bound of the covering radius. As example codebooks in $\PUN$, we consider the projective Pauli and Clifford groups, and the projective group of diagonal gates in the Clifford hierarchy, and find their minimum distances. We furthermore discuss single-qubit codebooks arising as products of the projective Clifford + $\mathbf{T}$ and projective Clifford + $\sqrt{\mathbf{T}}$ constructions in $\mathcal{PU}_2$, and obtain their minimum distance, distortion, and covering radius. Finally, we verify the analytical results by simulation. 
\end{abstract}

\begin{IEEEkeywords}
Projective unitary group, volume, kissing radius, Hamming bound, Gilbert-Varshamov bound, rate--distortion function.
\end{IEEEkeywords}

\section{Introduction}\label{section_Introduction}
In quantum computing, the design of quantum algorithms can be seen as
a decomposition of a unitary matrix using a set of universal gates. It
is well known that the set of Clifford gates combined with a
non-Clifford gate forms a set of universal gates for quantum
computation~\cite{nielsen2000quantum}. Exact decomposition or approximation of an arbitrary unitary matrix
using a set of universal gates has been addressed
in~\cite{matsumoto2008representation,kliuchnikov2015practical,
  selinger2012efficient}. In~\cite{matsumoto2008representation}, the
total number of single-qubit gates that can be represented by the
Clifford+T gates is calculated. An algorithm for finding a T-optimal
approximation of single-qubit Z-rotations using Clifford+T gates is
proposed in~\cite{kliuchnikov2015practical}, which is capable of
handling errors down to $10^{-15}$. Approximating an arbitrary
single-qubit gate from the special unitary group using Clifford+T
gates, up to any given error threshold, is proposed
in~\cite{selinger2012efficient}.

In quantum computation, the overall phase is irrelevant since it does
not affect the measurable properties of a quantum
system~\cite{nielsen2000quantum}. Hence, the gate approximation should be
considered in the projective unitary group $\PUN$ rather than in the
unitary group or the special unitary group. $\PUN$ consists of the
equivalence classes of $n \times n$ unitary operations that differ by
a global phase~\cite{boya2003volumes}. Understanding the geometry of the projective unitary group is thus fundamental for constructing reliable quantum gates and enabling universal quantum computation.

A  chordal distance, which is invariant to global phase rotations and thus suitable for $\PUN$, is considered in~\cite{kliuchnikov2015practical, A2011,
  mukhopadhyay2021composability, gheorghiu2022t}. In~\cite{A2011},
this metric is used for constructing the optimal fault-tolerant approximation of arbitrary gates with a set of discrete
universal gates. Using this metric, 
the error approximations of universal gates are discussed in
\cite{mukhopadhyay2021composability}.
Furthermore, the $\mathbf{T}$-count and $\mathbf{T}$-depth of any multi-qubit unitary,
which are crucial for optimizing quantum circuits, are analyzed in
\cite{gheorghiu2022t}.

The volume of a small ball is needed for deriving bounds of
packing and covering problems. The volume of a small ball in the
unitary group, Grassmannian, and Stiefel manifolds are
well understood~\cite{henkel2005sphere,
  barg2006bound, han2006unitary, dai2008quantization, AWTH2017}.
However, $\PUN$ remains largely unexplored in the literature,
particularly in terms of volume analysis and theoretical bounds. 

In~\cite{boya2003volumes}, volumes of $\UN$, $\SUN$ and $\PUN$ are computed. The volume of $\UN$, however, differs from the volume induced by natural geodesic and chordal distances~\cite{hua1959theory}, which yields meaningful coding bounds~\cite{AWTH2017}. For coding theoretical bounds it is essential for the volume of the space to be commensurate with the metric used to measure distances; the prefactors have to be correct, not only the scaling behavior. For this reason, we have to recompute the volume of $\PUN$.

 The kissing radius, analogous to the packing radius in linear codes \cite{sole1995packing}, plays a pivotal role in various applications, including the optimization of sphere-decoder algorithms \cite{schenk2009stopping, schenk2010stopping}. Also, the kissing radius relates to rate--distortion theory as it is the smallest possible distance from a codeword to the border of its Voronoi cell discussed in \cite{pitaval2013joint,mondal2007quantization}. Based on the volume of ball in the Grassmannian manifold, several bounds are derived for the rate–distortion tradeoff assuming that the cardinality of codebooks is sufficiently large \cite{dai2008quantization}.

\def\matT{\mathbf{T}}
Motivated by this, we consider the  chordal distance in $\PUN$ and compute the volume of a small ball. Using this volume,
we derive the Hamming upper and Gilbert-Varshamov (GV) lower bounds.
In addition, we obtain upper and lower bounds for the kissing radius
as a function of the minimum distance of the codebook in $\PUN$, and
establish a tight Hamming bound. We derive upper and lower bounds for
the distortion rate function. Furthermore, as examples of codebooks
in $\PUN$, we consider the projective Pauli group, the projective
Clifford group, and the group of projective diagonal gates in the
Clifford hierarchy, and determine their minimum distances. In the special case of ${\cal PU}_2$, we in addition consider codebooks arising from a product of a finite number of elements in the third and fourth level of the Clifford hierarchy, i.e. products of Clifford gates with $\matT$- or $\sqrt{\matT}$-gates, respectively. Finally,
through numerical results, we verify the validity of our analyses.

The rest of this paper is organized as follows:
Section~\ref{Sec:prilim} provides preliminaries. We derive the volume of metric balls for $\PUN$ in Section~\ref{Sec:3}, and give the
Hamming upper and GV lower bounds. 
Section~\ref{Sec:KissingRadiusandDistortionRate} provides the upper
and lower bounds for the kissing radius as a function of the minimum
distance. Also, we obtain bounds on the rate--distortion function, lower bound of covering radius and approximated value of covering radius in
 $\PUN$. Section~\ref{sec:6} discusses the simulation results, and Section~\ref{sec:8} concludes the paper.

\section{Preliminaries} \label{Sec:prilim}

\subsection{The Projective Unitary Group}

\def\rmH{{\rm H}}
The unitary group is the set of complex-valued matrices fulfilling the unitarity constraint,
\be
\UN=\{\mathbf{U} \in \mathbb{C}^{n\times n} ~|~ \mathbf{U}\mathbf{U}^\rmH =\mathbf{U}^\rmH\mathbf{U}=\mathbf{I}\}.
\ee
Here $(.)^\rmH$ denotes the Hermitian conjugate.
The unitary group is a  Lie group, and as such a differential manifold. The dimension  of $\UN$  as a real manifold is $\dim\, \, {\UN}=n^2.$
The center of $\UN$ is ${\cal U}_1$, any element of the form $e^{i \theta} \matI$ 
is in the center. 

An extrinsic distance between two elements 
$ \mathbf{U}$ and $ \mathbf{V}$ in $\UN$ is given by the Frobenius norm. 
In this paper, we shall use Euclidean distances based on a scaled Frobenius norm as
\begin{equation}
    d_F(\mathbf{U},\mathbf{V})=\frac1{\sqrt{2n}}\,\norm{\mathbf{U}-\mathbf{V}}_F=\sqrt{1-\frac1n \mathcal{R}\left[\Tr\left( \mathbf{U}^\rmH\mathbf{V} \right) \right]}\, .
    \label{eq:eucUN}
\end{equation}
For clarity, we define the scaling factor
\begin{equation}\label{Eq:scalingfactor}
  s_n = \frac1{\sqrt{2n}}
\end{equation}
which makes volume results more transparent. With this scaling, the distances are in $[0,\sqrt{2}]$ independently of $n$, with the maximum distance reached when $\mathbf{V} = - \mathbf{U}$.

The special unitary group $\SUN$ is defined as
\begin{equation}
\SUN = \{\, \mathbf{U} \in \UN \mid \det(\mathbf{U}) = 1 \,\}.
\end{equation}
$\UN$  is included in the semidirect product of $\SUN$  and $\calU_1$. The center of $\SUN$ is $\mathbb{Z}_n$, any $n$th root of unity $e^{2 \pi i\,m/n}$ for $m=0,\ldots, n-1$ times the identity is in the center. As $\SUN$ has a linear representation in terms of unitary matrices with determinant 1, an extrinsic distance on $\SUN$ is given directly by \eqref{eq:eucUN}. 

The projective unitary group $\PUN$ is a group of $n\times n$ complex
valued matrices which can be represented in the quotient geometry as
$\UN/\U1$. The dimension of $\PUN$ is $n^2-1$, and the elements are equivalence
classes $[\mathbf{U}] = \left\{ \alpha\matU \,\mid\, |\alpha|=1 \right\}$:
 \begin{equation}
 \PUN=\{ [\mathbf{ U}] \mid \mathbf{U}\in \mathcal{ U}_n\}\,,
 \label{eq:PUNdef}
 \end{equation}
which can be represented by any unitary matrix  $\alpha\mathbf{U}$ belonging to the class. For notational simplicity, we will denote  elements of $\PUN$ simply by a representative $\bf U$ of the class, when there is no possibility for confusion.

In this paper, we use the following extrinsic metric~\cite{A2011}:
\begin{equation}\label{Eq:PhaseInvDistMetric}
    d(\mathbf{U},\mathbf{V})=\sqrt{1-\frac{1}{n}\left\lvert \mathrm \Tr\left( \mathbf{U}^\rmH\mathbf{V} \right) \right\rvert},
\end{equation} 
for $\mathbf{U}, \mathbf{V} \in \PUN$. This is a metric on $\PUN$, as it does not depend
on the overall phase of the representation $\mathbf{U}$ of an element in $\PUN$. As this distance is not measured along the manifold, but along a chord that is cutting through the embedding space, following~\cite{conway1996packing} we call this the {\it chordal distance} on $\PUN$. We have normalized this distance such that $d(\matU,\matV)\leq 1$, with the maximal distance achieved when $\matU^\rmH\matV$ is traceless.

\subsection{Packing and Covering Problems}

A finite subset of
points in manifold $\mathcal{M}$
\begin{equation}
    \mathcal{C} = \{\mathbf{C}_1, \dots, \mathbf{C}_{\lvert \mathcal{C} \rvert}\} \subset \mathcal{M}, 
\end{equation}
is a $(\left| \mathcal{C} \right|, \delta)$-code, with
\begin{equation}
     \delta=\min\{d(\mathbf{C}_i, \mathbf{C}_j): \mathbf{C}_i, \mathbf{C}_j \in \mathcal{C}, i\neq j \}
 \end{equation}
the minimum distance.

The packing problem is to fit a maximal set of non-overlapping balls
of a given radius $R$ into the space. The centers of packing balls thus give rise to a code with $\delta\geq 2 R$.   
The standard Hamming bound is a packing bound that provides an upper
bound for the cardinality of a code given its minimum distance.

The covering problem is to find the minimum number of overlapping
balls of a given radius, required to cover the entire space. The Gilbert-Varshamov bound is a covering bound that provides a lower bound on the cardinality of the code, given its minimum distance. 

The GV lower bound and the Hamming upper bound in a manifold~\cite{henkel2005sphere, barg2002bounds} for the cardinality $|\mathcal{C}|$ are related 
to volumes of a metric ball as
$$
\underset{\text{Gilbert-Varshamov bound}}{\frac{1}{\mu( B(R))}}\leq|\mathcal{C}| \leq \underset{\text{ Hamming upper bound}}{\frac{1}{\mu( B(\frac{R}{2}))},}
$$
where $\mu$ is the measure of the metric ball in the manifold $\mathcal{M}$;
\begin{equation} \label{Eq:ballMeasure}
    \mu\left( B(R) \right) = \frac{{\rm Vol} \left( B(R) \right)}{{\rm Vol} (\mathcal{M})},
\end{equation}
where ${\rm Vol} \left( B(R) \right)$ is the volume of the ball with
radius $R$ in the manifold
satisfying $\mu(\mathcal{M})=1$.

In non-Euclidean geometry, the maximum radius of non-overlapping packing
balls, known as the kissing radius, may be larger than half of the
minimum distance $\delta$. The kissing radius may thus be used to enhance the Hamming bound. The kissing radius of a code $
\mathcal{C} $ is defined as
\begin{equation}
\varrho = \sup_{\substack{B_{ \mathbf{C}_l}(R) \cap B_{ \mathbf{C}_k}(R) = \emptyset \\ \forall (k, l), k \neq l}} R,    
\end{equation}
where 
\begin{equation}
B_{\mathbf{C}_i}(R) = \{\mathbf{P} \in \mathcal{M} \, : \, d(\mathbf{P}, \mathbf{C}_i) \leq R\}
\end{equation}
is the metric ball of radius $r$ centered on the codeword $\mathbf{C}_i$. 



\subsection{Codebooks in the Projective Unitary Group}

\def\ZZ{\mathbb{Z}}
For generic $n=2^m$ with $m\in\ZZ_+$, we shall consider three families of codebooks in $\PUN$: the projective Pauli group, the projective Clifford group, and the diagonal part of the Clifford Hierarchy. In addition, for $n=2$, we shall consider finite products of 3rd and 4th level Clifford hierarchy elements. These codebooks play a crucial role in quantum computation theory \cite{Calderbank1998,matsumoto2008representation,can2020kerdock, conway1985finite, cui2017diagonal, woit2017quantum}. 

\subsubsection{ Projective Pauli Group }
The  $2\times 2$ Pauli matrices are:
\[
\mathbf{X} = 
\begin{bmatrix}
0 & 1 \\
1 & 0
\end{bmatrix}, \quad
\mathbf{Y} = 
\begin{bmatrix}
0 & -i \\
i & 0
\end{bmatrix}, \quad
\mathbf{Z} = 
\begin{bmatrix}
1 & 0 \\
0 & -1
\end{bmatrix}.
\]
Higher dimensional Pauli matrices can be defined as 
\begin{equation*}\label{eqa5}
{\bf D}({\bf a}, {\bf b}) = 
\mathbf{X}^{a_1} \mathbf{Z}^{b_1} \otimes \mathbf{X}^{a_2} \mathbf{Z}^{b_2} \otimes \cdots \otimes \mathbf{X}^{a_m} \mathbf{Z}^{b_m}, 
\end{equation*}
with binary vectors
${\bf a} = \left[a_1, ..., a_m \right]^T, {\bf b} = \left[b_1,  ...,
  b_m \right]^T$. The $n= 2^m$-dimensional Pauli group is defined as~\cite{pllaha2020weyl}
\begin{equation}
    \mathcal{P}_n = \left\{ \kappa \mathbf{D}\left( \mathbf{a, b} \right) \, | \, \mathbf{a, b} \in \mathbb{F}_2^m,  \kappa \in \mathbb{Z}_4\right\},
\end{equation}
where
\begin{equation}\label{Eq:CenterofCliffordHierarch}
    \mathbb{Z}_k \triangleq \{ e^{ \frac{2\pi i}{k} q}
    \mid q = 0, 1, ..., k-1 \}.
\end{equation}
is the cyclic group with $k$ elements. We shall interchangeably use the multiplicative and additive representations of $\ZZ_k$, with the latter being the additive group $\ZZ/k\ZZ$ of integers modulo $k$.  

Also, we define Hermitian Pauli matrices as
\begin{equation}\label{Eq:HWmatrixes}
{\bf E}({\bf a}, {\bf b}) = i^{a^T b} {\bf D}({\bf a}, {\bf b}) \equiv {\bf E}({\bf c}), 
\end{equation}
where ${\bf c} = \left[ \mathbf{a}, \mathbf{b} \right] \in \mathbb{F}_2^{2m}$. We shall also use the normalized versions:
\begin{equation}
\tilde{{\bf E}}({\bf c }) = \frac{1}{\sqrt{n}} {\bf E }({\bf c}),
\end{equation}
which form an orthonormal basis of the $n^2$-dimensional complex vector space of $n \times n$ complex matrices.  The trace of these matrices satisfies:
\[
\text{Tr}({\bf E}({\bf c})) = 0, \quad \text{if } {\bf c} \neq 0,
\]
so that ${\bf E}({\bf 0}) = {\bf I}_n$ is only basis matrix with non-vanishing trace, where $\mathbf{I}_n$ denotes $n\times n$ identity matrix. 
The elements of the Pauli group either commute or anti-commute, and their product rule is:
\[
{\bf E}({\bf c}) {\bf E}({\bf c}') = \pm {\bf E}({\bf c}') {\bf E}({\bf c}) = \pm i {\bf E}({\bf c} + {\bf c}').
\]

The projective Pauli group is defined  as $\widetilde{\mathcal{P}}_n=\mathcal{P}_n/{\mathbb{Z}_4}
$.
Note that
$\widetilde{\mathcal{P}}_n$ has cardinality $n^2= 2^{2m}$.

\subsubsection{Projective Clifford Group}

The 
Clifford group is defined as~\cite{Calderbank1998}
\begin{equation}
    \mathcal{G}_n = \{ {\bf G} \in \mathcal{{U}}_n \mid {\bf G}^\rmH \mathcal{P}_n {\bf G}\subset \mathcal{P}_n \}.
    \label{eq:CliffDef}
\end{equation}
The group of unitary automorphisms of $\mathcal{P}_n$ is $\mathcal{G}_n$ and the group of inner automorphisms  is $\widetilde{\mathcal{P}}_n $. The unitary outer automorphism group is thus given by \cite{can2020kerdock, conway1985finite}
\begin{equation*}
\mathcal{G}_n / \widetilde{\mathcal{P}}_n \cong \text{Sp}(2m, 2),
\end{equation*} 
the binary symplectic group. This is the group of all binary $ 2m \times 2m $ matrices that fulfill:
\begin{equation*}
{\bf F} {\bf \Omega } {\bf F}^T = {\bf \Omega},
~~\text{where}~~
{\bm \Omega} = 
\begin{bmatrix}
{\bf 0}_m & {\bf I}_m \\
{\bf I}_m & {\bf 0}_m
\end{bmatrix}.
\end{equation*}
The isomorphism between the outer automorphisms and the symplectic group takes the form:
\begin{equation}\label{eqa24}
{\bf G_F}^\rmH {\bf  E}({\bf c}) {\bf G_F} = \pm {\bf E}({\bf c\, F}),
\end{equation}
i.e., for each symplectic binary matrix $ {\bf F} $, there exists a unitary transform $ {\bf G_F} $ which takes the Pauli element corresponding to the binary $2m$-dimensional vector $ {\bf c} $ to the element corresponding to $ {\bf c\, F}$, up to a sign. The sign is determined by the multiplications in $ \mathcal{P}_n $. There are $ 2^m $ binary degrees of freedom, corresponding to the inner automorphisms. Explicit details on this can be found in \cite{can2020kerdock}.
The identity element in the automorphism group corresponds to the identity in the symplectic group:
\begin{equation}
    {\bf G}_{{\bf I}_{2m}} ={\bf I}_n.
\end{equation}

According to \eqref{eq:CliffDef}, the center of the Clifford group would be $\calU_1$. However, conventionally the center is taken to be 
$\ZZ_8$, for which the Clifford group is linear group freely generated by Hadamard, Phase and CNOT gates~\cite{Calderbank1998}. The projective Clifford group is defined as $\tilde{\mathcal{G}}_n = \mathcal{G}_n / \mathbb{Z}_8$.
The cardinality of $\tilde{{\mathcal{G}}}_n$ is~\cite{pllaha2022binary}
\begin{equation}\label{CardinalityPC}
    \left| \tilde{\mathcal{G}}_n \right|
=2^{m^2+2m} \prod_{i=1}^{m} (2^{2i} - 1)\,.
\end{equation}


\subsubsection{Projective diagonal part of the Clifford hierarchy} 

As the first example of a codebook from a higher level of the Clifford hierarchy, we consider the  
diagonal Clifford hierarchy of level $k$ denoted by
$\mathcal{D}_{n,k}$, which forms a group. For $k < m$ it can be generated by the set of gates~\cite{Zheng2008} 
\[
\left\langle \mathbf{Z}_i\left[\frac{\pi}{2^k}\right], \bm{\Lambda}^1_{i_1,i_2}\left(\mathbf{Z}\left[\frac{\pi}{2^{k-1}}\right]\right), \dots, \bm{\Lambda}^{k-1}_{i_1,\dots,i_k}\left(\mathbf{Z}[\frac{\pi}{2}]\right) \right\rangle,
\]
where 
$\mathbf{Z}_j[\frac{\pi}{2^k}] = \exp\left(\frac{i\pi }{2^k}
\mathbf{Z}_j\right)$ are  $\frac{\pi}{2^k}$-rotations in the direction of $\mathbf{Z}_j$, the Pauli $\mathbf{Z}$
acting on the $j$th qubit, 
and $ \bm{\Lambda}^k(\mathbf{U}) $ denotes the $ k $-controlled $
\mathbf{U} $ gate, acting on $ k+1 $ qubits. Here, $ i $, $ i_1, i_2,
\dots, i_k $ run over all qubits. For $m\leq k$ a similar set of
generating gates truncated at $m'=m-1$ control gates construct
$\mathcal{D}_{n,k}$.
The projective diagonal part of the Clifford hierarchy is defined as
$\tilde{\mathcal{D}}_{n,k} = \mathcal{D}_{n,k} / \mathbb{Z}_{2^k}$. The cardinality of $\tilde{\mathcal{D}}_{n,k}$ is given by
\cite{anderson2024groups}
\begin{equation}\label{Eq:DPCH}
\lvert \tilde{\mathcal{D}}_{n,k} \rvert = \prod_{j=0}^{\min(k-1, m-1)} \left(2^{k-j}\right)^{\binom{m}{j+1}}.
\end{equation}

\subsubsection{Projective Semi-Clifford codebook} 

The semi-Clifford codebook $\mathcal{C}_{n,k}$  is defined as a collection of unitary matrices that can be expressed as $\mathbf{U} = \mathbf{G}_1 \,\mathbf{D}\, \mathbf{G}_2,$ where $\mathbf{G}_j \in \mathcal{G}_n$ and $\mathbf{D} \in \mathcal{D}_{n,k}$~\cite{Zheng2008}. Similarly, we can define the projective semi-Clifford  codebook $\tilde{\mathcal{C}}_{n,k}$ using the projective Clifford group and the projective diagonal part of the Clifford hierarchy. For single-qubit operations, the cardinality of the projective semi-Clifford codebooks can be computed as 
\begin{equation}
    \left|\tilde{\mathcal{C}}_{2,k}\right|=24\left( 3\cdot 2^{k-2}- 2 \right)
\end{equation}

\subsubsection{Codebooks of products of 3rd and 4th level Clifford hierarchy elements}

In~\cite{matsumoto2008representation} single-qubit gate operations were approximated using codebooks of products of Clifford+$\mathbf{T}$-gates, i.e. products of 3rd level Clifford hierarchy elements.  We define a codebook with the maximal number  $l$ of $\mathbf{T}$-gates in the codewords, interleaved with Clifford-group elements. The cardinality of the codebook considered in~\cite{matsumoto2008representation}  
is $192 \left( 3 \cdot 2^l-2 \right)$, where $192$ is the cardinality of the single-qubit Clifford group. We denote by 
$\tilde{\mathcal{T}}_l$ the restriction of this codebook to the projective unitary group $\mathcal{PU}_2$.   It follows that the cardinality of the codebook with at most $l$  $\mathbf{T}$-gates is 
\begin{equation}
    \lvert \tilde{\mathcal{T}}_l \rvert = 24 \left( 3 \cdot 2^l-2 \right).
\end{equation}

Motivated by this codebook, we also consider using a similar codebook with circuits of at most $l$ stages of $\mathbf{S} = \sqrt{\mathbf{T}}$ gates, which means that the codebook consists of products of elements in the $4$th level of the Clifford hierarchy. We denote this codebook by $\tilde{\mathcal{S}}_l $. Using a similar argument as in~\cite{matsumoto2008representation}, the cardinalities of these codebooks become
\begin{equation}
\lvert \tilde{\mathcal{S}}_l \rvert = 24 \left( \frac{9 \left( 6^l - 1 \right)}{5} + 1 \right).
\end{equation}



\section{Geometry of the Projective Unitary Group}\label{Sec:3}

In this section, we first discuss metrics on $\PUN$, $\SUN$ and $\UN$, and then find the corresponding volumes. Using the volume of $\PUN$, we derive a measure of a small metric ball in $\PUN$ with respect to metric given by~\eqref{Eq:PhaseInvDistMetric}. In addition, we provide the Hamming upper and GV lower bounds in $\PUN$.

\subsection{Distances on $\UN$, $\SUN$ and $\PUN$}
\noindent 
{\bf Distances in $\UN$:} 
As discussed in \cite{AWTH2017}, the unitary group $\UN$ forms a Lie group whose Lie algebra corresponds to its geodesic curves. Moreover, the Frobenius norm naturally reduces to an infinitesimal metric on the tangent space, defining the local geometric structure of $\UN$.

The relation between the extrinsic Euclidean distance \eqref{eq:eucUN} and the infinitesimal metric on $\UN$ can be found as follows~\cite{AWTH2017}. Defining a skew Hermitian matrix $\mathbf{A}$ in the tangent space of $\UN$ from $\mathbf{U}^\rmH\mathbf{V}= \exp(\mathbf {A})$, the 
geodesic between $\mathbf{U}$ and  $\mathbf{V}$ is given by $ \mathbf{U} \exp(t\mathbf {A})$ for $t\in[0,1]$. For an infinitesimal displacement we have  $\mathbf{V}
=\mathbf{U}\exp(d \mathbf{A})\approx\mathbf{U}(\mathbf{I}+d\mathbf{A})$, and the infinitesimal metric on $\UN$
\begin{eqnarray}
\norm{\mathbf{U}-\mathbf{U}(\mathbf{I}+d\mathbf{A})}_F
&=&\norm{\mathbf{U}d\mathbf{A}}_F =\norm{d\mathbf{A}}_F \cr
&=& \sqrt{\mathrm \Tr(d\mathbf{A}^\rmH  d\mathbf{A})}\,.
\label{Eq:infinitesimalmetric}
\end{eqnarray}
For clarity it is worth noting that the distance \eqref{eq:eucUN} we use is based on a scaled infinitesimal metric with scaling $s_n$.   

In addition to the extrinsic distance, there is a  natural intrinsic geodesic distance between $\matU$ and $\matV$, arising from this infinitesimal metric. 
Explicit forms of the  Euclidean and geodesic distances can be given in terms of principal angles.
%
%
Consider the eigenvalue decomposition
\begin{equation}\label{Eq:Decomposition}
 \mathbf{U}^\rmH  \mathbf{V}=  \mathbf{Q}\,
    \mathrm{diag}\!\left(e^{i\phi_1},\dots,e^{i\phi_n}\right)
    \mathbf{Q}^\rmH , 
\end{equation}
where $\mathbf{Q}$ is a unitary matrix.
The eigenvalues are of the form $e^{i {\phi_m}}$, and
\be
\boldsymbol{\phi}=(\phi_1, \phi_2, \dotsc, \phi_n )
\in[-\pi,\pi]^n
\label{eq:princeAngles}
\ee
is a vector of $n$ principal angles. 

The Euclidean distance \eqref{eq:eucUN} on $\UN$, expressed in terms of the principal angles becomes
\begin{equation}
    d_F(\mathbf{U},\mathbf{V})=\sqrt{\frac{2}{n}\sum_{i=1}^{n}\text{sin}^2\frac{\phi_i}{2}}
\end{equation}
while the geodesic distance is
\be
d_g(\matU,\matV) = \Vert \boldsymbol{\phi}\Vert_2\,.
\label{eq:geodUN}
\ee

\noindent 
{\bf Distances in $\SUN$:} 
The infinitesimal metric as well as the extrinsic and intrinsic distances are induced from the ones for $\UN$, i.e., \eqref{eq:eucUN} and \eqref{eq:geodUN}. The only difference is that as the exponential map is limited to traceless skew-Hermitian matrices, the
trace of the diagonal matrix of principal angles is 0, 
thus the principal angles are constrained  such that  $\sum_{i=1}^n \phi_i$ =0.

\noindent 
{\bf Distances in $\PUN$:} 
As $\PUN $ is a Lie group (see~\cite{hall2013quantum}), its geodesic can be described using its Lie algebra 
$$
\mathfrak{pu}(n) \cong \mathfrak{u}(n)/ \{ia \mathbf{I}\}=\{\mathbf{A}+ia\mathbf{I}: a\in \mathbb{R}, \mathbf{A}\in \mathfrak{u}(n) \}\,,
$$
where $\mathfrak{u}(n)$ is the Lie algebra of $\UN$. 
\def\matA{\mathbf{A}}
\def\matB{\mathbf{B}}
\def\RR{\mathbb{R}}
The infinitesimal metric of $\PUN$ is given by the induced Euclidean distance in  $\mathfrak{pu}(n)$. The Euclidean distance between the cosets represented by skew-Hermitian matrices $\matA$ and $\matB$ can be computed as a minimization over the cosets as
\begin{equation}
    \Vert \matA - \matB \Vert_{2,\mathfrak{pu}} = 
     \inf_{a\in \RR} \Vert \matA - \matB + ic\matI \Vert_{2}
\end{equation}
Writing $c = a -b + \epsilon$ with $a = \Tr(\matA)/n$ and $b = \Tr(\matB)/n$, we can define the traceless representatives 
$\widebar\matA  = \matA-a\matI$ and  $\widebar\matB  = \matB-b\matI$ in cosets $[\matA]$ and $[\matB]$, respectively. The squared argument of the infimum now becomes
\begin{eqnarray}
    \Vert \widebar\matA - \widebar\matB + i\epsilon \matI \Vert_{2}^2 = \Vert \widebar\matA - \widebar\matB \Vert_2^2 + 
    \epsilon^2 \Vert \matI \Vert_2^2\,.
\end{eqnarray}
The cross terms vanish as $\widebar\matA - \widebar\matB$ is traceless. This is minimized at $\epsilon = 0$, such that the induced Euclidean distance in the tangent space of $\PUN$ becomes 
\begin{equation}
    \Vert \matA - \matB \Vert_{2,\mathfrak{pu}} = 
    \Vert \widebar\matA - \widebar\matB\Vert_2\,.
    \label{eq:PUnTangentSpace}
\end{equation}

A geodesic equivalence class curve in $\PUN $ is given by
\begin{equation}\label{Eq:Geodesiccurve}
[\bm{\gamma}(t)] = \mathbf{U}\, e^{t(\mathbf{A}+ia\mathbf{I})}=\mathbf{U} e^{t\,[\mathbf{A}]},~~~~~~ 0\leq t \leq 1
\end{equation}
where $\mathbf{A}+ia\mathbf{I}$ is a skew-Hermitian matrix. Its  equivalence class $[\mathbf{A}] \in \mathfrak{pu}(n)$ such that
$[\bm{\gamma}(0)] = \mathbf{U}$, 
$[\bm{\gamma}(1)] = \mathbf{V}=\mathbf{U} \, e^{\mathbf{A}}$ and  we denote $[\mathbf{A}]=\mathbf{A}$. We have 
\begin{equation}\label{Eq:Geodesicdecomposition}
   e^{-i\theta}\mathbf{U}^\rmH  \mathbf{V} = e^{\mathbf{A}},~~\theta \in [0,2\pi)\,, 
\end{equation}
and as discussed in~\cite{FoundationFuncAnalysis} 
the geodesic distance between the equivalence classes $ [\mathbf{U}] $ and $ [\mathbf{V} ]$ in $ \PUN $ becomes
\begin{equation}\label{Eq:geodesicdist}
d_g([\mathbf{U}], [\mathbf{V}]) =  \inf_{\theta \in [0,2\pi)}\|e^{i\theta} \mathbf{U}^\rmH \mathbf{V} \|_2.
\end{equation}
The geodesic distance in $ \PUN $ can be expressed in terms of the principal angles  as 
\begin{equation}\label{Eq:geodesicdistinprin}
d_g([\mathbf{U}], [\mathbf{V}]) = \inf_{\theta \in [0,2\pi)}\big\| \lceil \boldsymbol{\phi}-\theta\rfloor 
\big\|_2\,,
\end{equation}
where we use the function 
\begin{equation}
    \lceil \chi \rfloor  = \chi - 2 \pi \lfloor \chi/2\pi + 1/2\rfloor
\end{equation}
which returns its arguments to the range $[-\pi,\pi]$, and used the natural definition of the difference of a vector and scalar.

The geodesic distance can be solved up to a search over a discrete set as follows. First note that without loss of generality, when computing the geodesic distance \eqref{Eq:geodesicdist}, the 
representatives in the equivalence classes $[\matU]$ and $[\matV]$ can be chosen to be special unitary matrices $\widebar\matU$ and $\widebar\matV$. The principal angles of $\widebar\matU^\rmH\widebar\matV$ are 
$\widebar\phi_m$ which fulfill $\sum_{m=1}^n \widebar\phi_m = 2 \pi k$ for $k\in\ZZ$. We have 

\begin{proposition}\label{propoPUN_SUN}
The geodesic distance on $\PUN$, induced by the geodesic distance \eqref{eq:geodUN} on $\UN$ is 
\begin{equation} \label{eq:geodPUNsun}
d_g([\matU],[\matV]) = \min_{k\in\ZZ_n}~\left\Vert
\left\lceil \boldsymbol{\bar\phi} + 2 \pi 
\frac{k}{n} \right\rfloor\right\Vert_2\,,
\end{equation}
where $\boldsymbol{\bar\phi}$ is a vector of principal angles $\bar\phi_n$ of  
$\widebar\matU^\rmH\widebar\matV$ for special unitary representatives $\widebar\matU$ and $\widebar\matV$ of $[\matU]$ and $[\matV]$.
\end{proposition}
\begin{IEEEproof}
Assume that the principal values $\phi_m$ are ordered in ascending order, and define the average principal angle as $\phi_{\rm ave} = \frac1{n} \sum_{m=1}^n\phi_m$. As norms are non-negative, the infimum over $\theta$ in \eqref{Eq:geodesicdistinprin} can be found from the squared norm. We thus define the objective function $f(\theta)  = \big\| \lceil \boldsymbol{\phi}-\boldsymbol{\theta}\rfloor 
\big\|_2^2$.  This is a continuous and periodic function in $[0,2\pi]$, and thus has a minimum. It 
is continuously differentiable except at values $\theta_m = \phi_m+\pi$, where $\lceil \phi_m- \theta\rfloor$ changes from $\phi_m-\theta$ to $\phi_m-\theta-2\pi$. For $\theta$ in the range $r_m = [\phi_m, \phi_{m+1})$, the objective function is 
\begin{equation}
    f_m(\theta) = \sum_{k=1}^m (\theta-\phi_k-2 \pi)^2 + \sum_{l=m+1}^n (\theta-\phi_l)^2\,.
\end{equation}
Due to periodicity, range $r_n$ can be defined to be $[\phi_n,2\pi) \cup [0,\phi_1)$. The objective function in range $r_m$ has a minimum at 
\begin{equation}
    \theta_m =  \phi_{\rm ave} + 2\pi\frac{m}{n}\,,
\end{equation}
which may or may not be in the range. As 
$$
\lim_{\theta\to\phi_m^-} \frac{df_{m-1}}{d\theta} = 2 n(\phi_m \!- \phi_{\rm ave}) - 4\pi (m-\!1)
= 
4\pi+\lim_{\theta\to\phi_m^+} \frac{df_{m}}{d\theta},
$$
the minimum cannot be at the boundary $\theta=\phi_m$ of two ranges $r_{m-1}$ and $r_{m}$. The minimum thus has to be within one of the regions, at one of the values $\theta_m$. For special unitary $\matU$ and $\matV$, we have $\phi_{\rm ave} =0$, thus the values $\theta_m$ reduce to $2\pi m/n$ for $m\in\ZZ_n$. 
\end{IEEEproof}

\def\calN{\mathcal{N}}
\def\matW{\mathbf{W}}
This metric is directly related to an alternative definition of $\PUN$ as the quotient of $\SUN$ with respect to its center $\mathbb{Z}_n$. Defining equivalence classes $[\widebar{\matU}] = \left\{e^{2 \pi i\,m/n}\, \widebar{\mathbf{U}} \mid m\in\mathbb{Z}_n\right\}$ in $\SUN$ we have 
 \begin{equation}
 \PUN=\{ [\widebar\matU] \mid \widebar{\mathbf{U}}\in \mathcal{SU}_n\}\,.
 \label{eq:PUNfromSUN}
 \end{equation}
Proposition \ref{propoPUN_SUN} shows that
this quotient representation and the conventional representation~\eqref{eq:PUNdef} are isometric. Moreover, $\PUN$ and $\SUN$ are isometric in the neighborhood of any point. For $\matU\in\SUN$, define the set
\begin{equation}
    \calN_{\matU} = \left\{ \matV ~\Big|~ 
    \matV\in\SUN: d_g(\matU,\matV) \leq \frac{\pi}{2n}  \right\}
\end{equation}
and the corresponding set of cosets
\begin{equation}
    \left[\calN_{\matU}\right] = \left\{\, [\matV]~ \big|~ \matV\in\calN_{\matU}\right\}\,.
\end{equation}
We now have

\begin{corollary}\label{coro:isom}
For any $\matU\in\SUN$, the neighborhoods $\calN_{\matU} \subset\SUN$
and $ \left[\calN_{\matU}\right] \subset\PUN$ are isometric.
\end{corollary}
\begin{IEEEproof}
By the triangle inequality, the the distance of any $\matV$ and 
$\matW\in\calN_{\matU}$ is $\leq \pi/n$. Thus each principal angle $\phi_k$ of 
$\matW^\rmH\matV$ is in the range $|\phi_k|<\pi/n$, and accordingly $\left|\lceil \phi_k + 2\pi m/n \rfloor\right| \geq |\phi_k|$ for all $m\in\ZZ_n$. According to Proposition~\ref{propoPUN_SUN}, the distance between$\matV$ and $\matW$ in $\SUN$ coincides with the distance between $[\matV]$ and $[\matW]$ in $\PUN$. 
\end{IEEEproof}

This shows that locally $\PUN$ and $\SUN$ are isometric, and confirms that the infinitesimal metrics of $\PUN$ and $\SUN$ coincide, as indicated by \eqref{eq:PUnTangentSpace}. 

Next consider the metric induced on $\PUN$ by the extrinsic chordal distance  \eqref{eq:eucUN} on $\UN$.  
For two equivalence classes $[\mathbf{U}], [\mathbf{V}] \in \PUN$, we define \cite{haah2023query}
\begin{eqnarray}
d([\mathbf{U}],[\mathbf{V}]) &=& \min_{\theta \in [0, 2\pi)} s_n \| \mathbf{U} - e^{i\theta} \mathbf{V} \|_F \cr
& =&  \sqrt{1-\frac1n 
\max_{\theta \in [0,2\pi)}
\mathcal{R}\left\{e^{i\theta}\Tr(\mathbf{U}^\rmH \mathbf{V})\right\}}\,. \nonumber
\end{eqnarray}
As for any complex number, ${\cal R}(z)\leq |z|$, with equality when the phase angle vanishes, it follows that this metric is equal to the  chordal distance \eqref{Eq:PhaseInvDistMetric}.

Writing this in terms of the principal angles \eqref{eq:princeAngles}, we get
\begin{eqnarray}
    \!\! d(\matU,\matV) &\!=&\!\sqrt{1-\frac{1}{n}\Big|\sum\nolimits_{ j=1 }^{n}e^{i\phi_j}\Big|}\cr
     &\!=&\!\sqrt{1-\frac1n \sqrt{n+2\sum\nolimits_{i< j} \cos(\phi_i-\phi_j)}}\,
      \label{Eq:Principalangledistance}
\end{eqnarray}
which also explicitly shows global phase invariance.


{\bf Operator norm and trace distance:}
In \cite{nielsen2000quantum}, the relationship between the operator norm 
$d_{\rm O}(\mathbf{U},\mathbf{V})={\max_\psi}
\norm{(\mathbf{U}-\mathbf{V})|\,\bm{\psi}\rangle},$ where the maximum is over all pure states $|\bm{\psi}\rangle$, and the trace distance 
\begin{eqnarray*}
    d_{\rm Tr}(\matU,\matV) &=& \mathrm{Tr}\left(\sqrt{(\mathbf{U} - \mathbf{V})^\rmH (\mathbf{U} -  \mathbf{V}) }\right)\cr
&=& \sqrt{2}\sum_{m=1}^n \sqrt{1-\cos\phi_m}
\end{eqnarray*}
for single-qubit rotations is discussed, in the context of approximating unitary operators. In determining these distances, the global phase of the unitary matrices play a significant role. For example, for both of these metrics, the
distance between $\matU$ and $-\matU$ is maximal, while for
\eqref{Eq:PhaseInvDistMetric}, their distance is zero, as they come from the same equivalence class. If these metrics are used for finding  approximations of quantum operators, undue attention is paid to global phases.

\textbf{Diamond distance on $\PUN$:}
The diamond distance, given by the diamond norm, is used to measure distances between quantum channels~\cite{Kitaev2002}. In \cite{kliuchnikov2023shorter}, the diamond distance
$d_{\diamond}(\mathbf{U},\mathbf{V}) = \| \mathbf{U} - \mathbf{V} \|_{\diamond}$
between unitary matrices $\mathbf{U}$ and $\mathbf{V}$ was discussed. It is given by
the diameter of the smallest disk in the complex plane containing all eigenvalues of $\mathbf{U}^{\rmH}\mathbf{V}$. This distance is clearly global phase invariant, and has a maximum value of 2, the diameter of the unit circle.

From \eqref{Eq:Decomposition}, the eigenvalues 
of $\mathbf{U}^{H}\mathbf{V}$ are
$\{ e^{i\phi_1}, \dots, e^{i\phi_n} \}$. The distance between two eigenvalues is then
\begin{equation}\label{Eq:Euclicdeandis}
    |e^{i\phi_i}-e^{i\phi_j}|=2\Big|\sin\Big(\frac{\phi_i-\phi_j}{2}\Big)\Big|=2\sin\Big(\frac{\Delta_{i,j}}{2}\Big),
    \end{equation}  
where $\Delta_{i,j}= \min \{
\left|\phi_i-\phi_j\right|, 2 \pi - \left|\phi_i-\phi_j\right|\}
$, with the understanding that angle differences are measured modulo $2\pi$. As 
a consequence, $\Delta_{i,j}\in[0, \pi]$. 

When $d_\diamond(\mathbf{U},\mathbf{V})<2$, there is a disk with radius $r<1$ which covers all the eigenvalues. The maximum length of an arc of the unit circle, covered by a disk with $r<1$ centered anywhere, is $<\pi$. Thus iff  $d_\diamond(\mathbf{U},\mathbf{V})<2$, all eigenvalues of $\mathbf{U}^{H}\mathbf{V}$ are in a half plane. 
Without loss of generality, in this case we can rotate the global phase such that all principal angles are in the half space $|\phi_i|\leq \pi/2$. Defining
\begin{equation}\label{Eq:phasedifference}
   \omega= \max_{i\neq j} \Delta_{i,j}
\end{equation}
we then have 
\begin{equation}\label{Eq:diamondnorm}
d_{\diamond}(\mathbf{U},\mathbf{V}) 
= 
2\,\max_{i\neq j} \sin\Big(\frac{\Delta_{i,j}}{2}\Big)
= 2\,\sin\Big(\frac\omega{2}\Big).
\end{equation}

For $d_\diamond(\mathbf{U},\mathbf{V})<2$
one has the following inequalities between the diamond and chordal distances, achieved by concrete eigenphase placements:

\textit{Upper bound}, given by symmetric eigenvalue split. For diamond norm \eqref{Eq:diamondnorm} with $\omega <\pi$, all principal angles $\phi_i$ are within an arc of the unit circle with length $\omega$. The maximum chordal distance is found by taking half of the principal angles with value $\omega/2$, half with $-\omega/2$. Therefore, 
we have  $\Big|\sum\limits_{ k=1 }^{n}e^{i\phi_k}\Big|=\frac{n}{2}\Big|e^{i\,\omega/2}+e^{i\,\omega/2}\Big|$, and from \eqref{Eq:Principalangledistance} we get 
\begin{equation}\label{Eq:phasedis}
 d(\mathbf{U},\mathbf{V})=\sqrt{1-\cos(\omega/2)}\,.
\end{equation} 

\textit{Lower bound} given by two extreme phases, with others zero. The minimal chordal distance is found by putting  two phases at $\pm \Phi/2$ and the remaining $n-2$ phases at $0$. leading to
$$
\Big|\sum\limits_{ k=1 }^{n}e^{i\theta_k}\Big|
=\Big|e^{i\omega/2}+e^{i\omega/2}+n-2\Big|= n-2+2\cos(\omega/2)\,.
$$ 
From \eqref{Eq:Principalangledistance} we then have 
\begin{equation}
    d(\mathbf{U},\mathbf{V})=\sqrt{\frac{2}{n}}\sqrt{1-\cos(\omega/2)}
\end{equation}

Hence the chordal distance $d(\mathbf{U},\mathbf{V})$ is bounded as:
  \begin{equation}
      \sqrt{\frac{2}{n}}\sqrt{1-\cos(\omega/2)} \leq d(\mathbf{U},\mathbf{V}) \leq \sqrt{1-\cos(\omega/2)}
  \end{equation}
From \eqref{Eq:diamondnorm} we obtain $\cos(\omega/2)=\sqrt{4-\frac1{4}d_{\diamond}^2}$, leading to bounds on the chordal distance in terms of the diamond distance:
\begin{equation}\label{Eq:bound}
      \sqrt{\frac{2}{n}}\sqrt{1-\sqrt{1-
      \frac1{4}d_{\diamond}^2}} \leq d(\mathbf{U},\mathbf{V}) \leq \sqrt{1-\sqrt{1-\frac1{4}d_{\diamond}^2}}
  \end{equation}
Correspondingly, we have the following bounds on the diamond distance given the chordal distance $d$: 
\begin{equation}\label{Eq:bound2}
     2 \,d \sqrt{2-d^2}\leq d_\diamond(\mathbf{U},\mathbf{V}) \leq 
     d \sqrt{n \left(4-n\,d^2\right)}\,,
\end{equation}
with the upper bound valid for $d\leq \sqrt{2/n}$, where the bound reaches the value 2.  

When $\omega$ is small we have $\sqrt{1-\frac1{4}d_{\diamond}^2}\approx 1-\frac{d_{\diamond}^2}{8}$. However, in the Taylor series of the bounds \eqref{Eq:bound}, any truncation reduces the value of the function. Truncation can thus be done in the lower bound, but strictly speaking not in the upper bound. In the upper bound, the coefficient of the quadratic term has to be increased to have a valid bound in a non-zero range. From \eqref{Eq:bound} we thus get 
\begin{equation}\label{Eq:bound1}
      \frac{1}{2\sqrt{n}}\, d_{\diamond}(\mathbf{U},\mathbf{V}) 
      \leq d(\mathbf{U},\mathbf{V}) 
      \leq \frac{1}{\sqrt{a}}\,d_{\diamond}(\mathbf{U},\mathbf{V}) 
  \end{equation}
where $a<8$ is a constant, and the upper bound is valid in the range $d_\diamond <  \frac12\,\sqrt{8 a - a^2}$.
For small values of the chordal distance, 
from \eqref{Eq:bound2} we correspondingly get the diamond distance bounds 
\begin{equation}\label{Eq:bound3}
      \sqrt{a}\, d(\mathbf{U},\mathbf{V}) 
      \leq d_{\diamond}(\mathbf{U},\mathbf{V}) \leq 2\sqrt{n}\, d(\mathbf{U},\mathbf{V}) \,
  \end{equation}
where $a<8$, the lower bound is valid in the range $d\leq \frac12 \sqrt{8 - a}$, and the upper bound for $d\leq \sqrt{2/n}$. 

It is straight forward that similar bounding relationships could be derived between the geodesic and diamond distances. 

In particular, when $n=2$ the upper and lower bounds coincide. For $d_\diamond < 2$, there is an one-to-one relationship between the chordal and diamond distances. These is also a one-to-one relationship to the geodesic distance. From (\ref{eq:geodPUNsun},\ref{Eq:Principalangledistance},\ref{Eq:diamondnorm}), when $n=2$ and the 
principal angle difference $\omega$ is small,
these distances read
\begin{eqnarray}
    d_g(\matU,\matV) &=& \frac1{\sqrt{2}}\,\omega\\
    d(\matU,\matV) &=& \sqrt{1-\cos\left(\frac{\omega}2\right)}\\
    d_\diamond(\matU,\matV) &=& 2\sin\left(\frac{\omega}2\right)\,,
\end{eqnarray}
and for infinitesimal $\omega$ we have 
\begin{equation}
    \frac12 d_g(\mathbf{U},\mathbf{V}) \simeq  d(\mathbf{U},\mathbf{V})
    \simeq \frac1{\sqrt{8}}\, d_{\diamond}(\mathbf{U},\mathbf{V})\,.
    \label{eq:infinitesimal2}
\end{equation}

The analysis above shows that packing and covering results for small balls, derived for the chordal distance~\eqref{Eq:PhaseInvDistMetric} are pertinent for the diamond distance as well, and the same holds for the geodesic distance. In particular, for $n=2$ there is a one-to-one correspondence between the chordal, diamond and geodesic distances, and all results for the chordal distance can be directly mapped to results for the diamond or geodesic distance. Below, we concentrate on the chordal distance, due to its analytic tractability.

\subsection{Volume of Metric Balls in the Projective Unitary Group}

\subsubsection{Infinitesimal Metric Scaling}

In Riemannian geometry, the volume of a space, or a subset of a space is obtained by integrating the infinitesimal metric, which depends on the notion of infinitesimal displacement. When considering measures of balls, or packing and covering bounds in Riemannian manifolds, it is paramount that the distance metric used to characterize a packing or a covering, such as minimum distance, covering radius, or distance to the closest approximant, is measured in a manner which is commensurate with measuring volumes in the manifold, i.e. with the infinitesimal form of the geodesic distance. 

To proceed with volume computations we first verify that a scaled version of the geodesic distance $d_g $ in $\PUN$  
and the phase-invariant metric $d$ are the same in an infinitesimal limit, i.e., that \eqref{eq:infinitesimal2} generalizes to arbitrary $n$. We have:


\begin{proposition} \label{propo:infinte}
The  chordal distance $d$ in $\PUN$ coincides with $s_n = 1/\sqrt{2n}$ times the geodesic distance $d_g $ in the infinitesimal case.
\end{proposition}
\begin{IEEEproof}
From Proposition~\ref{propoPUN_SUN} and Corollary~\ref{coro:isom} it follows that when elements $[\matU]$ and $[\matV]\in\PUN$ are infinitesimally close, their distance coincide with the distance of their $\SUN$ representatives $\widebar\matU$ and $\widebar\matV$.
We thus have  $\widebar\matV = \widebar\matU \exp(d\mathbf {A})
\approx\mathbf{U}(\mathbf{I}+d\mathbf{A})$,
where $d\matA$ is an infinitesimally small traceless skew-Hermitian matrix. The infinitesimal geodesic distance is measured in the tangent space as
\begin{eqnarray}
d_g(\widebar\matU,\widebar\matV)
&\approx& 
\norm{\widebar\matU-\widebar\matU(\mathbf{I}+d\mathbf{A})}_F
=\norm{\widebar\matU d\mathbf{A}}_F \cr
&=&   \left(
\Tr(d\mathbf{A}^\rmH d\mathbf{A})
\right)^{\frac{1}{2}} \label{Eq:Geodesicapp}
\end{eqnarray}
To compute the corresponding phase-invariant distance, we take the corresponding classes in $\PUN$ represented by special unitary matrices $\matU$ and $\matV$ and thus ignoring global phase. 
We consider
$$
\widebar\matU^\rmH \widebar\matV =\exp(\mathbf{A})\approx \mathbf{I}+d\mathbf{A} + \frac{1}{2}d\mathbf{A}^2\,.
$$
As $d\mathbf{A}'$ is traceless and infinitesimal, we get 
\begin{align*}
  \lvert \mathrm\Tr\left( \widebar\matU^\rmH \widebar\matV \right) \rvert &\approx\lvert \mathrm n+0-\frac{1}{2}\Tr\left(  d\mathbf{A}^\rmH d\mathbf{A} \right) \rvert \\&=
     n-\frac{1}{2}\Tr\left( d\mathbf{A}^\rmH d\mathbf{A} \right)\,, 
  \end{align*}
  and using \eqref{Eq:PhaseInvDistMetric} leads to
  \begin{equation}\label{Eq:Distanceapp}
      d(\mathbf{U},\widebar\matV)\approx \frac1{\sqrt{2n}}(\mathrm \Tr(d\mathbf{A}^\rmH d\mathbf{A}))^{\frac{1}{2}}\, 
  \end{equation}
  which is a  \eqref{Eq:Geodesicapp} up to a scaling with $s_n$. The statement follows directly.
\end{IEEEproof}

\subsubsection{Measure of Metric Balls}

According to Proposition \ref{propo:infinte}, to get proper packing and covering bounds for the  chordal distance  \eqref{Eq:PhaseInvDistMetric}, the volume of the $\PUN$ group manifold has to be computed with respective to a scaled infinitesimal metric, which is commensurate with \eqref{Eq:PhaseInvDistMetric}.

The Euclidean $(D-1)$-sphere of radius $R$ in $\mathbb{R}^D$ is
defined as
$\mathcal{S}^{D-1}\left( R \right) = \big\{ \mathbf{x}\in
\mathbb{R}^D \, \big\vert \, \lVert \mathbf{x} \rVert_2 = R
\big\}$.
The volume of $\mathcal S^{D-1}(R)$ is given by
 \begin{equation}\label{Eq:Volumeofsphare}
      {\rm V}_D(R) = \frac{\pi^{D/2}}{\Gamma\left( \frac{D}{2} + 1 \right)} R^D.
\end{equation}

The volume of the unitary group $\UN$ computed with the inifinitesimal metric \eqref{Eq:infinitesimalmetric}
 is~\cite{hua1959theory} 
\begin{equation}\label{Eq:VolumeofUnitarygroup}
{\rm {Vol}}(\UN)=(2\pi)^{\frac{n(n+1)}{2}} \prod_{k=1}^{n-1}{\frac{1}{k!}}.
\end{equation}
For the volume of $\PUN$ we have:

\begin{theorem}\label{Theo:GlobalVoumeofPUN}
    The volume of the $\PUN$, commensurate with the chordal distance~\eqref{Eq:PhaseInvDistMetric} is 
  \begin{equation}\label{Eq:volumeofPUN}
    {\rm {Vol}}\left( \PUN \right)=2^{\frac{n-1}{2}}\pi^{\frac{n^2+n-2}{2}}n^{\frac{-n^2}{2}} \prod_{k=1}^{n-1}{\frac{1}{k!}}.
\end{equation}
\end{theorem}
\begin{IEEEproof}
    According to~\cite{boya2003volumes},
the volume of a homogeneous quotient space  $\mathcal{G}/\mathcal{K}$ arising
from the free and proper action of subgroup $\cal K$ on group $\cal G$
is ${\rm Vol} \left( \mathcal{G}\right)/{\rm
  Vol}\left( \mathcal{K} \right)$.
  
The volume of the $\UN$ with respect to the infinitesimal form of chordal distance~\eqref{Eq:PhaseInvDistMetric} is 
  \begin{equation}\label{Eq:ValumeofUN}
{\rm {Vol}}(\UN)=\frac{(2\pi)^{\frac{n(n+1)}{2}}}{(2n)^{\frac{n^2}{2}}}\prod_{k=1}^{n-1}{\frac{1}{k!}}.
\end{equation}
This follows directly from ~\cite{hua1959theory} by adding the  scaling factor \eqref{Eq:scalingfactor} to the volume element of the unitary group. 
 
It is important to understand that the subgroup forming the cosets in \eqref{eq:PUNdef} is isomorphic to ${\mathcal{U}_1}$, but {\it not} isometric. To find a metric on this subgroup, consider 
$\mathbf{X} = e^{i\theta} \mathbf{I}_n$ and $\mathbf{X}' = \mathbf{X}+d\mathbf{X}$ where $d\mathbf{X} = ie^{i\theta} \mathbf{I}_n \,d\theta$. The infinitesimal distance is given by
\begin{equation}
(ds)^2_{\mathcal{U}_1}=s_n^2\norm{\mathbf{X}-\mathbf{X}'}^2_F=s_n^2\norm{d\mathbf{X}}^2_F=s_n^2 n\,d^2\theta.
\end{equation}
Therefore, the subgroup divided away is isometric to a circle with radius $\sqrt{n}$, and the volume of subgroup $\mathcal{K}$ in $\UN$ is 
 $  {\rm {Vol}}(\mathcal{K}) = s_n\int_0^{2\pi} \sqrt{n} d\theta = \sqrt{2}\pi$. The statement follows directly.
\end{IEEEproof}
Similar to the proof of Theorem \ref{Theo:GlobalVoumeofPUN}, the volume of $\PUN$ with respect to the Frobenius norm without scaling is
\begin{equation}\label{Eq:PUNvolume}
    {\rm {Vol}}\left( \PUN \right)=\frac{(2\pi)^{\frac{n(n+1)}{2}}}{2\pi\sqrt{n}}\prod_{k=1}^{n-1}{\frac{1}{k!}}.
\end{equation}

For the measure of the metric ball in  $\PUN$ we have 
\begin{corollary}\label{Cor:Measurofball}
As $R \to 0$, the measure of a metric ball $B(R)$ in $\PUN$ with respect to the chordal distance~\eqref{Eq:PhaseInvDistMetric} or the geodesic distance~\eqref{eq:geodPUNsun} scaled with $s_n$ is
\begin{equation}\label{Eq:Measurofball}
    \mu(B(R))=c_{n}\,R^{D}(1+\mathcal{O}(R^2)) 
\end{equation}
    where $D=n^2-1$ is the dimension of $\PUN$, and 
    $$
    c_{n}=\frac{(2\pi)^{-\frac{(n-1)}{2}}n^{\frac{n^2}{2}}}{\Gamma(\frac{n^2-1}{2}+1)}\prod_{k=1}^{n-1}k!\,.
    $$
\end{corollary}
\begin{IEEEproof}
The measure of metric ball in $\PUN$ with respect to the metric~\eqref{Eq:PhaseInvDistMetric}, or~\eqref{eq:geodPUNsun} scaled with $s_n$, can be written as 
\begin{equation}\label{Eq:BallMeasurRel}
    F_d\left( R \right)= \Pr\{ d \le R \} = \mu\left( B\left( R \right) \right).
\end{equation}
 The volume of a
small ball can be well approximated by the volume of a ball of equal
radius in the tangent space~\cite{AWTH2017} as
\begin{equation}\label{Eq:VolBr}
    {\rm Vol}\left( B(R) \right) = {\rm V}_D(R)\left( 1 + O(R^2)\right).
\end{equation}
Substituting~\eqref{Eq:VolBr} and~\eqref{Eq:volumeofPUN}in~\eqref{Eq:ballMeasure} and considering~\eqref{Eq:BallMeasurRel} completes the proof.
\end{IEEEproof}

To the best of our knowledge, in the literature no previous work has provided expressions of the volumes of $\SUN$ and $\PUN$ which are exact enough for considering packing and covering bounds. Several studies have addressed the computation of these volumes~\cite{tilma2002generalized, tilma2004generalized, bertini2006euler, marinov1981correction, boya2003volumes}. However in these works the volumes of the groups and subgroups are measured such that the geometry of embedding small balls in the manifolds are not properly captured.

Since $\PUN = \SUN / \ZZ_n$, it follows from \eqref{Eq:PUNvolume} that
\begin{equation}
 {\rm Vol(\SUN)} = n~{\rm Vol(\PUN)}.
 \end{equation}
Using this form for $\SUN$ and \eqref{Eq:volumeofPUN} for $\PUN$ gives more accurate Hamming and Gilbert-Varshamov bounds on these manifolds than using the results from the literature.  

It is worth to note that $\PUN$ is not isometric with the Complex Stiefel manifold ${\mathcal{V}}_{{n,n-1}}^{\mathbb{C}}$, discussed in \cite{krishnamachari2008volume}. The Stiefel manifold can be interpreted in two ways, either as the quotient space ${\mathcal{V}}_{{n,p}}^{\mathbb{C}} \cong {\mathcal{{U}}_n}/{\mathcal{{U}}_{n-p}}$, where $\mathcal{U}_{n} $ is the unitary group, or as the space of rectangular $n \times  (n-p)$ matrices. The former interpretation gives rise to the canonical metric discussed in \cite{edelman1998geometry}.

To understand the difference of the geometries of the quotient space interpretation of the Stiefel manifold for $p=n-1$ and $\PUN$, a detailed discussion is needed. In the quotient geometry, a point in ${\mathcal{V}}_{{n,p}}^{\mathbb{C}}$ is an equivalence class of unitary matrices
\begin{equation}\label{Eq:Stiefel}
[\mathbf{U}] = \Bigg\{\mathbf{U} \begin{pmatrix}
\mathbf{I}_p & \mathbf{O} \\
\mathbf{O} & \bm{\Upsilon}
\end{pmatrix}: \quad \bm{\Upsilon} \in\mathcal{{U}}_{n-p} \Bigg\}.
\end{equation}
Note that $\dim\, \, {\mathcal{V}}_{{n,p}}^{\mathbb{C}}=\dim\,  \mathcal{{U}}_{n}-\dim\,  \mathcal{{U}}_{n-p}=2np-p^2 $. As discussed in \cite{AWTH2017}, for $\mathbf{U}, \matV \in \mathcal{V}_{n,p}^{\mathbb{C}},$  the geodesic distance between $\mathbf{U}$ and  $\mathbf{V}= \mathbf{U} \exp(\mathbf{U}^\rmH  \bm{\Delta})$  is
\begin{equation}\label{eqq1}
    d_g(\mathbf{U},\mathbf{V})= \norm {\bm{\Delta}}_F=( \norm{\mathbf{A}}_F^2+2\norm{\mathbf{B}}_F^2)^{\frac{1}{2}},
\end{equation}
where $\bm{\Delta} =\mathbf{U}\begin{pmatrix}
 \mathbf{A} & -\mathbf{B}^\rmH \\
 \mathbf{B} &  \mathbf{O} 
\end{pmatrix}$ with $\mathbf{A}^\rmH =-\mathbf{A}$ and $\mathbf{B} \in \mathbb{C}^{(n-p)\times p}.$

From  \eqref{Eq:Stiefel}, an element of the Stiefel manifold ${\mathcal{V}}_{{n,n-1}}^{\mathbb{C}}$ is an equivalence class 
$$
\left\{
\mathbf{U} \begin{pmatrix}
\mathbf{I}_{n-1} & \mathbf{O} \\
\mathbf{O} & e^{i\theta}
\end{pmatrix}: \quad e^{i\theta} \in \mathcal{U}_{1}
\right\}
$$
while an element of $\PUN$ is
$
\left\{
\left(e^{i\theta}\,\mathbb{I}_n\right) \, \mathbf{U}: \quad e^{i\theta}\,\mathbb{I}_n \in \mathcal{U}_{1}
\right\}.
$
In both cases, cosets of a $\calU_1$ subgroup are considered. The geometry of these two, w.r.t to the 
extrinsic chordal distance is different, however. For ${\cal V}_{n,n-1}$, the volume of the 
$\calU_1$-component is $2\pi$ , and we have 
${\rm Vol}\left(\mathcal{V}_{n,n-1}\right) = \mathcal{V}_{n,n-1}/2\pi$, as discussed in \cite{krishnamachari2008volume}.
That is, for any value of $n$, ${\rm Vol}\left({\mathcal{V}}_{n,n-1}^{\mathbb{C}}\right)$ depends only on ${\rm Vol}(\UN)$, whereas the volume of $\PUN$ in~\eqref{Eq:PUNvolume} depends on both ${\rm Vol}(\UN)$ and the additional factor $\sqrt{n}$.

\section{Minimum-distance Bounds on 
$\PUN$}\label{Sec:KissingRadiusandDistortionRate}

The GV and Hamming bounds provide lower and upper bounds on the cardinality of a codebook in the manifold~\cite{henkel2005sphere}. 
In the following, we provide these bounds for  $\PUN$.
There exists a codebook $\mathcal{C}$ in $\PUN$ with cardinality $\lvert \mathcal{C} \rvert$ and the minimum distance $\delta$ with respect to the metric ~\eqref{Eq:PhaseInvDistMetric} such that 
\begin{equation}\label{Eq:GVBound}
    \frac{1}{\mu(B(\delta))} \leq\lvert \mathcal{C} \rvert.
\end{equation}
Also,  for any $(\left| \mathcal{C} \right|, \delta)$-codebook in  $\PUN$
\begin{equation}\label{Eq:HammBound}
    \left| \mathcal{C} \right|\leq\frac{1}{\mu(B(\frac{\delta}{2}))}.
\end{equation}
The Gilbert-Varhamov bound arises from a covering argument. If $|{\calC}|$ balls 
$B(\delta)$ do not cover the manifold, there is room to add one more point which is at least at distance $\delta$ from all other points.

The Hamming bound is a packing bound, literally bounding the number of codewords surrounded by $B(\frac{\delta}{2}))$-balls that can be packed into the manifold. The Hamming bound can be enhanced by analyzing the kissing radius. 

\subsection{Kissing  Radius Bounds of  Projective Unitary Group}\label{sec:4}
In this section, we derive upper and lower bounds for the kissing
radius $\varrho$ as a function of the minimum distance of a code in
$\PUN$ with respect to the  chordal distance. The kissing radius is relevant in cases where packings and coverings are sparse. When the curvature of the manifold cause, e.g., minimum chordal distances $\delta$ of packing balls to differ from twice the distances to a geodesic midpoint, the kissing radius becomes relevant. 

Moreover, we establish a tight
Hamming bound in this context, using the  density
$\Delta(\mathcal{C})$ of a code.
For a code with cardinality $K$ and kissing radius $\varrho$, the
density is \cite{AWTH2017}:
$$
\Delta(\mathcal{C}) =K\mu( B(\varrho))\,.
$$  

We define the {\it midpoint} between two points on a manifold with respect to a distance metric to be the point which is at equal distance from the endpoints, such that the distance from the two points is minimal. 

\def\matM{\mathbf{M}}
\def\matQ{\mathbf{Q}}
\def\matL{\mathbf{L}}
\def\matOme{\mathbf{\Omega}}
\begin{lemma}\label{Lem:MidPoint}
Let $\mathbf{U}, \mathbf{V} \in \PUN$. The midpoint of $\matU$ and $\matV$ w.r.t  the
metric~\eqref{Eq:PhaseInvDistMetric}, is given by the geodesic midpoint
  \begin{equation}\label{Eq:midPoint}
      \mathbf{M} = \mathbf{U\, \Omega}\, \sqrt{\mathbf{L}}\, \mathbf{\Omega}^\rmH = \mathbf{V\, \Omega}\, \sqrt{\mathbf{L}^\rmH}\, \mathbf{\Omega}^\rmH
      \,,
  \end{equation}
  where $\mathbf{W}=\mathbf{U}^\rmH \mathbf{V}
  = \mathbf{\Omega L}\mathbf{\Omega}^\rmH $ with $\mathbf{L} = \diag\left(e^{i\phi_1}, ..., e^{i\phi_n} \right)$.
 \end{lemma}
 \begin{IEEEproof}
The shortest path between two points 
$\matU$ and 
$\matV$ on $\PUN$, as measured by the geodesic distance~\eqref{eq:geodPUNsun}, is described by the geodesic curve~\eqref{Eq:Geodesiccurve}. Hence we have
$\bm{\gamma}(0) = \mathbf{U}, \bm{\gamma}(1) = \mathbf{V}=\mathbf{U} e^{\mathbf{A}}$. 
The geodesic midpoint $\mathbf{M}$ is given by $\bm{\gamma}(1/2)$ which can be written in the form of the first equality of~\eqref{Eq:midPoint}. The second equality follows by using $\matU = \matV\matW^\rmH$ in the first form.
The squared chordal distance between $\mathbf{U}$ and $\mathbf{M}$ in $\PUN$ is
\begin{align*}
d^2(\mathbf{U},\mathbf{M}) &= 1 - \frac1n \Big| \mathrm{Tr}(\mathbf{U}^\rmH  \mathbf{M}) \Big| 
     = 1 - \frac1n \Big| \mathrm{Tr}\,\sqrt{\matL} \Big| \,.
    \end{align*}
Using the second form of $\mathbf{M}$ in~\eqref{Eq:midPoint}, we have  
\begin{align*}
d^2(\mathbf{V},\mathbf{M}) &= 1 - \frac1n \Big| \mathrm{Tr}(\mathbf{V}^\rmH  \mathbf{M}) \Big| 
    = 1 - \frac1n \Big| \mathrm{Tr}\,\sqrt{\matL^\rmH} \Big| \,.
\end{align*}
It follows that
$ d\left( \mathbf{U}, \mathbf{M} \right) = d\left( \mathbf{V}, \mathbf{M} \right)$.

Thus the geodesic midpoint is at the same distance from the end points w.r.t. the chordal distance. As the geodesic is defined with \eqref{eq:geodPUNsun}, and we measure distances using \eqref{Eq:Principalangledistance},
there is a logical possibility, however, that some other point $\matM'\in \PUN$ exists, which is not on the geodesic, and fulfills $ d\left( \mathbf{U}, \mathbf{M}' \right) = d\left( \mathbf{V}, \mathbf{M}' \right) <  d\left( \mathbf{V}, \mathbf{M} \right)$.
Without loss of generality we may represent such a point as  $\matM' = \matM\matQ$ for some unitary $\matQ$. Considering the trace-part in the distance \eqref{Eq:Principalangledistance}, we then should have 
$$
\left|\Tr  \sqrt{\matL} \,\matOme^\rmH\matQ\matOme\right| = 
\left|\Tr  \sqrt{\matL^\rmH}\, \matOme^\rmH\matQ\matOme\right| > 
\left|\Tr  \sqrt{\matL}\right|\,.
$$
From the equality it follows that all diagonal elements in $\matOme^\rmH\matQ\matOme$ should be real. As the principal angles $\phi_j$ are in $[-\pi,\pi]$, the square roots $e^{i\phi_j/2}$ of the eigenvalues lie in the half-plane with non-negative real part. Accordingly, the trace is maximized for $\matOme^\rmH\matQ\matOme =\matI$, and thus no such $\matM'$ exists. 
 \end{IEEEproof}
\vspace{2mm}

The kissing radius of a given code is hard to determine since it depends on the minimum distance of the code and principal angles between codewords~\cite{pitaval2011density}. The minimum distance $\delta$ gives rise to a constraint on the principal angles between two points on the manifold at distance $\delta$. The kissing radius $\varrho$ is then given as the distance to the midpoint according to Lemma~~\ref{Lem:MidPoint}. Bounds on the kissing radius are found by maximizing and minimizing $\varrho$ given the minimum distance constraint:
%
%
\begin{equation}\label{Eq:Principalangle}
\varrho=\sqrt{1-\frac{1}{n}\Big|\sum\limits_{ j=1 }^{n}e^{\frac{i\phi_j}{2}}\Big|}
~~\text{such that}~~
n(1-\delta^2)=\Big|\sum\limits_{ j=1 }^{n}e^{i\phi_j}\Big|\,.
\end{equation}
Let $\bar{\varrho}$ be an upper bound  and  $\underline{\varrho}$ be a lower bound on the kissing radius $\varrho$. Then, the following theorem gives bounds for the kissing radius: 

\begin{theorem}\label{Theo:Boundonthekissingradius}
For any code $(\left| \mathcal{C} \right|, \delta)\in \PUN$, the kissing radius $\varrho$ is bounded as
$$\underline{\varrho} \leq \varrho\leq \bar{\varrho},$$ where
$\underline{\varrho}=\sqrt{1-\sqrt{1-\frac{\delta^2}{2}}}$ and $\bar{\varrho}=\sqrt{1-\sqrt{ \frac{1+(1-\delta^2)^2}{2}}}$.
The corresponding  bounds on codebook density are
\begin{equation}\label{Eq:Boundoncodedensity}
     \left| \mathcal{C} \right|\mu(B(\underline{\varrho}))\leq \Delta(\mathcal{C})\leq \min\{1, \left| \mathcal{C} \right|\mu(B(\bar{\varrho}))\},.
\end{equation}
\end{theorem}
\begin{IEEEproof} 
According to \eqref{Eq:Principalangle}, we consider optimizing the kissing radius given the minimum distance. First we calculate upper bound of kissing  radius in the $\PUN$. We can see that
\begin{equation*}
    \Big|\sum\limits_{ j=1 }^{n}e^{i\phi_j}\Big|^2=n+ 2\sum\limits_{ 1\leq i < j \leq n }^{}\cos(\phi_i -\phi_j),
    \end{equation*} and
    \begin{equation}\label{Eq:inquality}
 \begin{split}
    \Big|\sum\limits_{ j=1 }^{n}e^{\frac{i\phi_j}{2}}\Big|^2
    &= n+ 2\sum\limits_{ 1\leq i < j \leq n }^{}\cos\Big(\frac{\phi_i -\phi_j}{2}\Big)\\
    &=n+ 2\sum\limits_{ 1\leq i < j \leq n }^{}\sqrt{\frac{\cos(\phi_i -\phi_j)+1}{2}}.\\
    \end{split} 
    \end{equation}
    For  $0\leq x \leq 1 $, we know~~  $\sqrt{x}\geq x.$ From this, it follows that
    \begin{multline*}
    n+ 2\sum\limits_{ 1\leq i < j \leq n }^{}\sqrt{\frac{\cos(\phi_i -\phi_j)+1}{2}}\\ \geq n+ 2\sum\limits_{ 1\leq i < j \leq n }^{}\bigg(\frac{\cos(\phi_i -\phi_j)+1}{2}\bigg)\\
     =\frac{n^2+\Big|\sum\limits_{ j=1 }^{n}e^{i\phi_j}\Big|^2}{2}
     =\frac{n^2+n^2(1-\delta^2)^2}{2}
     \end{multline*}
     From \eqref{Eq:inquality}, we have an inequality
     \begin{equation*}
         \Big|\sum\limits_{ j=1 }^{n}e^{\frac{i\phi_j}{2}}\Big|
    \geq {\sqrt{\frac{n^2+n^2(1-\delta^2)^2}{2}}}.
    \end{equation*}
Hence upper bound of kissing radius is: 
$$\varrho \leq \sqrt{1-\sqrt{ \frac{1+(1-\delta^2)^2}{2}}}.$$
For lower bound, we need to find
$\max \Big|\sum\limits_{ j=1 }^{n}e^{\frac{i\phi_j}{2}}\Big|$ such that $n(1-\delta^2)=\Big|\sum\limits_{ j=1 }^{n}e^{i\phi_j}\Big|.$

For $\max \Big|\sum\limits_{ j=1 }^{n}e^{\frac{i\phi_j}{2}}\Big|$ the triangle inequality states that
\begin{equation*}
\left|\sum_{j=1}^n e^{\frac{i\phi_j}{2}}\right| \leq \sum_{j=1}^n \left|e^{\frac{i\phi_j}{2}}\right| = n,
\end{equation*}
with equality if and only if all the vectors $e^{\frac{i\phi_j}{2}}$ are aligned, i.e., their arguments $\frac{\phi_j}{2}$ differ by a multiple of $2\pi$.
Similarly, the constraint $\left|\sum_{j=1}^n e^{i\phi_j}\right| = n(1 - \delta^2)$ implies that the vectors $e^{i\phi_j}$ are not fully aligned unless $\delta = 0$.

To maintain the constraint, we split the $n$ angles into two groups. We assume $\frac{n}{2}-k$,   $\phi_j$ would be same $\theta$ in one group and $\frac{n}{2}+k$,  $\phi_j$  would be same $\phi$ in other group.  These angles satisfy the constraint 
\begin{align*}
n(1 - \delta^2)=\left|\sum_{j=1}^n e^{i\phi_j}\right|
&=\left|\left(\frac{n}{2}-k\right)e^{i(\theta-\phi)}+\left(\frac{n}{2}+k\right)\right|
\end{align*}
i.e.,
\begin{align*}
    \left(n(1 - \delta^2)\right)^2&=\frac{2n^2}{4}+2k^2+2\left(\frac{n^2}{4}-k^2\right)\cos(\theta-\phi)
\end{align*}
\begin{equation*}
\frac{1}{2\left(\frac{n^2}{4}-k^2\right)}\left(\left(n(1 - \delta^2)\right)^2-\frac{2n^2}{4}-2k^2\right) =\cos(\theta-\phi).
\end{equation*}
So we have
\begin{equation*}
-1\leq\frac{\left(n(1 - \delta^2)\right)^2-\frac{2n^2}{4}-2k^2}{2\left(\frac{n^2}{4}-k^2\right)}\leq 1
\end{equation*}
if and only if $k=0.$ Without loss of generality we set
\begin{equation*}
\phi_1=\phi_2= \cdots = \phi_{\frac{n}{2}} =\theta ~~\text{and}~~ \phi_{\frac{n}{2}+1}=\phi_{\frac{n}{2}+2}= \cdots = \phi_{n} =\phi .
\end{equation*}
It follows that
$n(1-\delta^2)=\Big|\sum\limits_{ j=1 }^{n}e^{i\phi_j}\Big|=\Big|\frac{n}{2}e^{i\theta}+\frac{n}{2}e^{i\phi}\Big|,$
i.e., $4(1-\delta^2)^2=2+2\cos{(\theta-\phi)}.$ This implies that 
\begin{equation*}
\cos{(\theta-\phi)}= 2(1-\delta^2)^2-1.
\end{equation*}
Let us now examine, 
\begin{align*}
\Big|\sum\limits_{ j=1 }^{n}e^{\frac{i\phi_j}{2}}\Big|^2&=\Big|\frac{n}{2}e^{i\frac{\theta}{2}}+\frac{n}{2}e^{i\frac{\phi}{2}}\Big|^2
\\&=\frac{n^2}{4}\left(2+2\sqrt{\frac{\cos(\theta -\phi)+1}{2}}\right)
\\&=\frac{n^2}{4}\left(4-2\delta^2)\right).
\end{align*}
So,
 $\Big|\sum\limits_{ j=1 }^{n}e^{\frac{i\phi_j}{2}}\Big|=n\sqrt{1-\frac{\delta^2}{2}}.$
 Thus, we obtain the lower bound for the kissing radius:
 \begin{equation*}
     \varrho \geq \sqrt{1-\sqrt{1-\frac{\delta^2}{2}}}.
     \end{equation*}
\end{IEEEproof}

One of the central problems in coding theory is determining the maximum size of a codebook for a given minimum distance. 
Using the normalized volume of the metric ball $\mu( B(r))$, as given in Corollary \ref{Cor:Measurofball}, and applying Theorem \ref{Theo:Boundonthekissingradius}, we obtain the following refined version of the Hamming bound:

\begin{corollary}\label{Cor:Lowerboundofkissingradius}
      For any $(\left| \mathcal{C} \right|, \delta)$-code in $\PUN$, we have
      \begin{equation}\label{6.33}
       \lvert \mathcal{C} \rvert \leq \frac{1}{\mu( B(\underline{\varrho}))},
\end{equation}
where $\underline{\varrho}$ is given in Theorem~\ref{Theo:Boundonthekissingradius}.
\end{corollary}

\begin{IEEEproof}
From \eqref{Eq:Boundoncodedensity},
$ \left| \mathcal{C} \right|\mu(B(\underline{\varrho}))\leq \Delta(\mathcal{C})\leq 1$, and  we have   $\frac{\delta}{2} \leq \underline{\varrho}  $. It implies that \begin{equation*}
|\mathcal{C}| \leq\frac{1}{\mu( B(\underline{\varrho}))} \leq \frac{1}{\mu( B(\frac{\delta}{2}))}.
\end{equation*}

\end{IEEEproof}

\subsection{Minimum Distances of the Projective Unitary Codebooks}\label{sec:5}
The minimum distance of quantum codebooks plays a pivotal role in
facilitating both error correction and error detection
\cite{knill1997theory, pollatsek2001quantum,djordjevic2021quantum }.
Here, we find the minimum distance of the example codebooks in $\PUN$.

\begin{lemma}\label{Lemma:Tracesum}
For any $ n \times n $ matrix ${\bf{M}} $ and an orthonormal basis $ \{ {\bf{B}}^{(n)} \} $ for the vector space $ \mathcal{M}_\mathbb{C}(n) $, we have
\[
\sum_n \mathrm{Tr} \left({\bf{M}}^{H} {\bf{B}}^{(n)H} {\bf {M}} {\bf{B}}^{(n)} \right) = |\mathrm{Tr}\left( {\bf{M}})\right|^2.
\]
\end{lemma}
\begin{IEEEproof}
This follows directly from the completeness of the basis. The linear mapping from $ \mathcal{M}_\mathbb{C}(n) $ to itself, given by the sum of the outer products of the basis elements with themselves, is the identity. Explicitly, this means
\[
\sum_n b_{i,j}^{(n)*} b_{k,l}^{(n)} = \delta_{i,k} \delta_{j,l},
\]
where $ b_{i,j}^{(n)} $ are the matrix elements of $ {\bf B}^{(n)} $, and $ \delta_{i,k} $ is the Kronecker delta function.
\end{IEEEproof}
\begin{lemma}\label{Lemma:Kernal}
The inner products of the transformed Pauli matrices with themselves take the values
\[
\mathrm{Tr} \left({\bf  G}_{\bf F}^{H} \tilde{{\bf E}}({\bf c}) {\bf G}_{\bf F} \tilde{{\bf E}}({\bf c}) \right) =
\begin{cases} 
\pm 1 & \text{if } {\bf F({\bf c})} + {\bf c} = 0 \mod 2, \\
0 & \text{otherwise}.
\end{cases}
\]
\end{lemma}
\begin{IEEEproof}
This follows from \eqref{eqa24}, and the definition \eqref{Eq:HWmatrixes} of the Pauli matrices. From these we see that the matrix 
within the trace is 
${\bf E}({\bf F({\bf c})} + {\bf c}) / n,$ up to an integer power of 
$i$. The trace is non-vanishing only if this {\bf E} is proportional 
to identity. This occurs if and only if ${\bf F({\bf c})} + {\bf c}=0\mod 2 $, and according to \eqref{Eq:HWmatrixes}  the integer power of $ i $ determining the sign $ \pm 1 $.
\end{IEEEproof}
\begin{proposition}\label{Pop:minimumdistance}
   The minimum distances of $\widetilde{\mathcal{P}}_n$ and  $\tilde
   {\mathcal{G}}_n$ considering the chordal distance  are
    \begin{equation}
     \delta_p=1\,, \qquad \delta_c=\sqrt{1-\frac{1}{\sqrt{2}}}\,,
    \end{equation}
    respectively.
\end{proposition}
\begin{IEEEproof}
As $\widetilde{\mathcal{P}}_n$ is a group under multiplication, let
$\mathbf{U}^\rmH \mathbf{V}=\mathbf{W}=e^{ \frac{2\pi i}{2^k} q}{\bf D}({\bf a},{\bf b})
\in \widetilde{\mathcal{P}}_n$, i.e.,
$d(\mathbf{U},\mathbf{V})=d(\mathbf{I},\mathbf{W})$. Then 
$ \lvert \mathrm \Tr\left( \mathbf{I}^\rmH \mathbf{W} \right) \rvert  =\lvert \mathrm e^{ \frac{2\pi i}{2^k} q}\Tr\left({\bf D}({\bf a},{\bf b}\right)\rvert.$ We have

\begin{equation}\label{equation 8}
	   \lvert \mathrm e^{ \frac{2\pi i}{2^k} q}\Tr\left({\bf D}({\bf a},{\bf b}\right)\rvert = \begin{cases}
     n,  ~~~ {\bf a}={\bf b}\\
     0 ~~~~~ \text{otherwise}
    \end{cases}
	\end{equation}
From \eqref{Eq:PhaseInvDistMetric} and \eqref{equation 8}, the distance of any two codewords in  $\widetilde{\mathcal{P}}_n$ is zero or $1.$ Hence, $\delta_p = 1$.

For finding $\delta_c$, first we observe that 
\begin{equation}
  \lvert\mathrm{Tr}({\bf G}_{\bf F}) \rvert \leq \sqrt{\sum_{\bf c} \Big \lvert \mathrm{Tr} \left( {\bf G}_{\bf F}^{H} \tilde{{\bf E}}({\bf c}) {\bf G}_{\bf F} \tilde{{\bf E}}({\bf c}) \right) \Big\rvert}.
\end{equation}
This follows from Lemma \ref{Lemma:Tracesum}, as the matrices $ \tilde{{\bf E}}({\bf c}) $ form an orthonormal basis in $ \mathcal{M}_\mathbb{C}(n) $, and from the triangle inequality.
Using Lemma \ref{Lemma:Kernal}, a term in the sum over $ {\bf c} $ contributes a factor of 1 if $ {\bf F}({\bf c}) + {\bf c} = 0 \mod 2 $. The sum is over all binary $2m $-vectors, so the result is given by the number of vectors in the null space of $\bf{F} + \bf{I}_{2m} $, which evaluates to
\begin{equation*}
n_0 = 2^{2m - \mathrm{rank}(\bf{F} + \bf{I}_{2m})}.
\end{equation*}
For $\mathbf{F} = \mathbf{I}_{2m} $, we have $ n_0 = n^2 $, corresponding to $ \bf{G}_{\bf{F}} = \mathbf{I}_n $. Otherwise, for $ {\bf G}_{\bf F} \neq {\bf I}_n $, $ \mathrm{rank}(\bf{F} + \bf{I}_{2m}) \geq 1 $, so $n_0 \leq 2^{2m-1}$.
From the above inequality, we find
$|\mathrm{Tr}({\bf G}_{\bf F})| \leq \frac{n}{\sqrt{2}}.$
Since the bound depends only on $ |{\bf G}_{\bf F}| $, considering the center of $\tilde{\mathcal{G}}_n $ does not change the result. Using the equation \eqref{Eq:PhaseInvDistMetric} the minimum chordal distance between two matrices  in the $\tilde{\mathcal{G}}_n$ is
\begin{equation*}
\delta^2_c \geq \left(1 - \frac{2n}{n\sqrt{2}}\right).
\end{equation*}
Hence $ \delta_c=\sqrt{1-\frac{1}{\sqrt{2}}}.$

\end{IEEEproof}
\begin{proposition}\label{Pop:minimumdistanceofCHDG}
    The minimum distance of $\tilde{\mathcal{D}}_{n,k}$ with respect
    to~\eqref{Eq:PhaseInvDistMetric} is 
    
    \begin{equation*}
        \delta_d=\sqrt{1- \cos\left(\frac{\psi_k}{2}\right)},~ \text{where}~ \psi_k=\frac{2\pi}{2^k}.
    \end{equation*}
\end{proposition}
\begin{IEEEproof}
As $\tilde{\mathcal{D}}_{n,k}$ is a group under multiplication, let
$\mathbf{U}^\rmH \mathbf{V}=\mathbf{W}
\in \tilde{\mathcal{D}}_{n,k}$, i.e.,
$d(\mathbf{U},\mathbf{V})=d(\mathbf{I},\mathbf{W})$. Then, using a similar approach as in~\cite{bayanifar2023extended}, we can find the minimum distance. Basically, in an $m$-qubit system, setting $\mathbf{W} = \mathbf{I}^{\otimes (m-1)} \otimes \mathbf{Z}_{m}\left[ \frac{\pi}{2^k} \right]$ in $d\left( \mathbf{I}, \mathbf{W} \right)$ results in the minimum distance.
 \end{IEEEproof}
 \vspace{2mm}

In general, determining the minimum distances of the semi-Clifford, $\tilde{\mathcal{T}}_l $, and $\tilde{\mathcal{S}}_l $ codebooks is nontrivial; hence, we employ numerical simulations to estimate their minimum distances.

\section{Bounds on Codebook Distortion} 

In universal quantum computation, the goal is to approximate a given unitary gate by the closest element of a universal gate set. With a finite computational codebook, this inevitably leads to distortion---the executed circuit is only an approximation of the desired circuit. Accordingly, the average distortion, or the largest distortion may be of more interest for quantum computation than the minimum distances of the codebooks. 

\subsection{Rate--Distortion Function in $\PUN$}\label{sec:7}

The quantization problem of approximating the target using the codebook of available circuits, is directly related to rate--distortion theory.
The distortion rate function is defined as~\cite{dai2008quantization}
\begin{equation}
    \mathcal{D}^*(K)=\underset{\mathcal{C}:|\mathcal{C}| = K}\inf \mathcal{D}(\mathcal{C}),
\end{equation}
 where 
 \begin{equation}\label{Eq:Distortion}
     \mathcal{D}(\mathcal{C}) = \mathbb{E} \left[ \underset{\mathbf{P} \in \mathcal{C}}\min~d^2(\mathbf{P}, \mathbf{Q}) \right],
 \end{equation}
 where $\mathcal{C}$ in $\PUN$ with cardinality $\lvert \mathcal{C} \rvert = K$. Here, $\mathbf{Q}$ is an arbitrary point in the space.

Based on the volume of $\PUN$ given in Corollary~\ref{Cor:Measurofball},  the rate--distortion tradeoff is characterized by establishing lower and upper bounds on the rate--distortion function.
For a codebook $\mathcal{C}$ with sufficiently large cardinality $K$,
the rate--distortion function over the $\PUN$, with the chordal distance, can be bounded as
\begin{equation} \label{Eq:Distortionbound}
    \frac{D}{D + 2}\left(c_{n} K\right)^{-\frac{2}{D}} \leq \mathcal{D}^*(K) \leq \frac{2\Gamma\left( \frac{2}{D} \right)}{D} \left(c_{n} K\right)^{-\frac{2}{D}} \big( 1 + o(1) \big),
\end{equation}
where $D=n^2-1$ and
$c_{n}$ given in Corollary \ref{Cor:Measurofball}. This is an extension of the results in~\cite{dai2008quantization} for Grassmannian manifold to $\PUN$.

 As discussed in~\cite{dai2008quantization}, for a code $\left(K,
\delta \right)$ in $\PUN$ the distortion is upper bounded as
 \begin{equation}\label{Eq:SimpleDistBound}
     \mathcal{D}(\mathcal{C})\leq \left(\frac{\delta^2}{4}-1\right)K\mu(B(\delta/2))+1\,.
 \end{equation}
Note that in a flat space, the packing radius is $\frac{\delta}{2}$.
However,  in a non-flat geometry
$\frac{\delta}{2}\leq \varrho \leq \delta $. Therefore, for any code $(\left| \mathcal{C} \right|, \delta)\in \PUN$ and using the lower bound of the kissing radius, we can have a tighter upper bound on the distortion than~\eqref{Eq:SimpleDistBound}. Hence using~\eqref{Eq:SimpleDistBound} and Propositions \ref{Pop:minimumdistance} and \ref{Pop:minimumdistanceofCHDG}, we can obtain distortion upper bounds for codebooks $\widetilde{\mathcal{P}}_n$, $\tilde {\mathcal{G}}_n$ and $\tilde{\mathcal{D}}_{n,k}$, $\tilde{\mathcal{C}}_{2,k}$, $\tilde{\mathcal{T}}_l $ and $\tilde{\mathcal{S}}_l $.

\subsection{Covering radius}

The worst-case distortion is governed by the covering radius of the codebook in $\PUN$. With  $\mathcal{C} = \{\mathbf{C}_1, \dots , \mathbf{C}_K\}$ a codebook of 
$K$ 
points over $\PUN$.
The covering radius $\rho$ is
\begin{equation}\label{Eq:Coviringradius}
\rho=\max_{\mathbf{U} \in \PUN}\min_{1\leq i\leq K } d(\mathbf{C}_i, \mathbf{U}),\,
\end{equation}
and it is thus the square root of the maximum distortion. 

A lower bound for the covering radius follows directly from a covering argument. As $\PUN$ is a compact manifold, each open cover  of $\PUN$ has a finite sub-cover. 
Consider a ball $B_{\rho}(\mathbf{P}_i)$ centered at each codeword $\mathbf{P}_i \in \mathcal{C}$. By definition of the covering radius we have
$\PUN \subseteq \bigcup_{i=1}^{K} B_{\rho}(\mathbf{P}_i).$
      This implies that ${\rm {Vol}}(\mathcal{M})\leq K {\rm {Vol}}( B(\rho))$,
      i.e., $K\mu(B(\rho))\geq 1.$
 Using Corollary \ref{Cor:Measurofball}, 
 a lower bound of the covering radius is
\begin{equation}\label{Eq:Lowerboundofcoveringradius}
          \rho \geq \left(\frac{1}{c_{n}\,K}\right)^{1/D}.
 \end{equation}

The expected value of the covering radius of a random codebook consisting of $K$ points selected uniformly at random from the Grassmannian manifold can be found in \cite{breger2018points}, while a  general proof for the expected value of the covering radius of random codebooks on  compact manifolds is provided in  \cite{reznikov2016covering}. 

From these works, we find the expected value of the covering radius of a random codebook on  $\PUN$ with a sufficiently large cardinality  $K$ as follows:
Let ${\cal C}_K = \{\mathbf{P}_1, \dots , \mathbf{P}_K\}$ be a set of $K$ points selected independently and  uniformly at random from 
$\PUN$ with respect to the measure $\mu$. Then 
\begin{equation}
   \lim_{K \to \infty}\mathbb{E}\left(\rho\right)\left(\frac{K}{\log K}\right)^{\frac{1}{D}}=\left(\frac{{\rm {Vol}}(\PUN)}{\text{V}_D(1)}\right)^{\frac{1}{D}}.
\end{equation}
The covering radius can be approximated as
\begin{equation}\label{Eq:Approximatedcoveringradius}
\rho\approx \left(\frac{{\rm {Vol}}(\PUN)}{\text{V}_D(1)}\frac{\log K}{K}\right)^{\frac{1}{D}}.
\end{equation}

 \begin{figure}
     \centering
     \includegraphics[width=\linewidth]{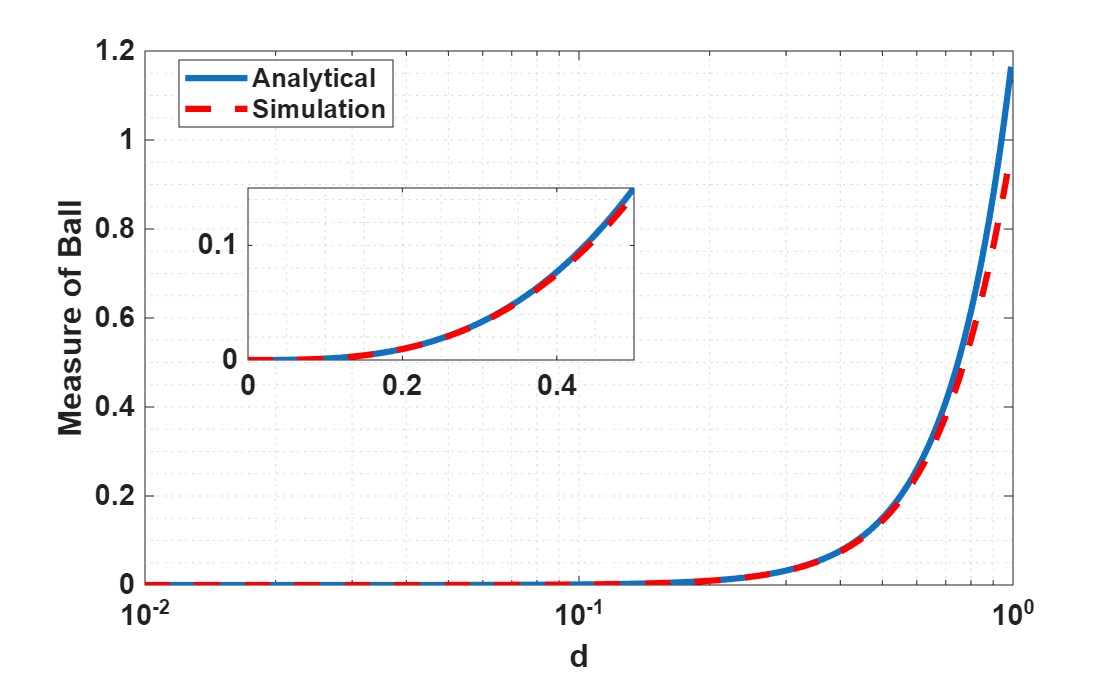}
     \caption{
     Theoretical and simulation results comparison of the measure of the ball in $\mathcal{PU}_2$, given by Corollary~\ref{Cor:Measurofball}.
     }
     \label{fig:00}
 \end{figure}

\section{Simulation Results}\label{sec:6}


 In this section, we verify the correctness of our analyses in $\PUN$ using numerical results. First, in Fig.~\ref{fig:00}, we consider the measure of the ball in $\PUN$ given by 
Corollary~\ref{Cor:Measurofball}. This figure illustrates the small
ball volume in $\PUN$ for $n=2$ in terms of the chordal
distance. The simulation results are obtained by averaging over $10^8$
unitary matrices generated uniformly at random with the Haar measure,
following~\cite{FrancescoRandMatGen2007}. Note that due to the
quotient structure, this also provides the Haar measure in $\PUN$. The
simulations results for small values of the distance matches with the
theoretical evaluation.

\begin{figure}
     \centering
     \includegraphics[width=\linewidth]{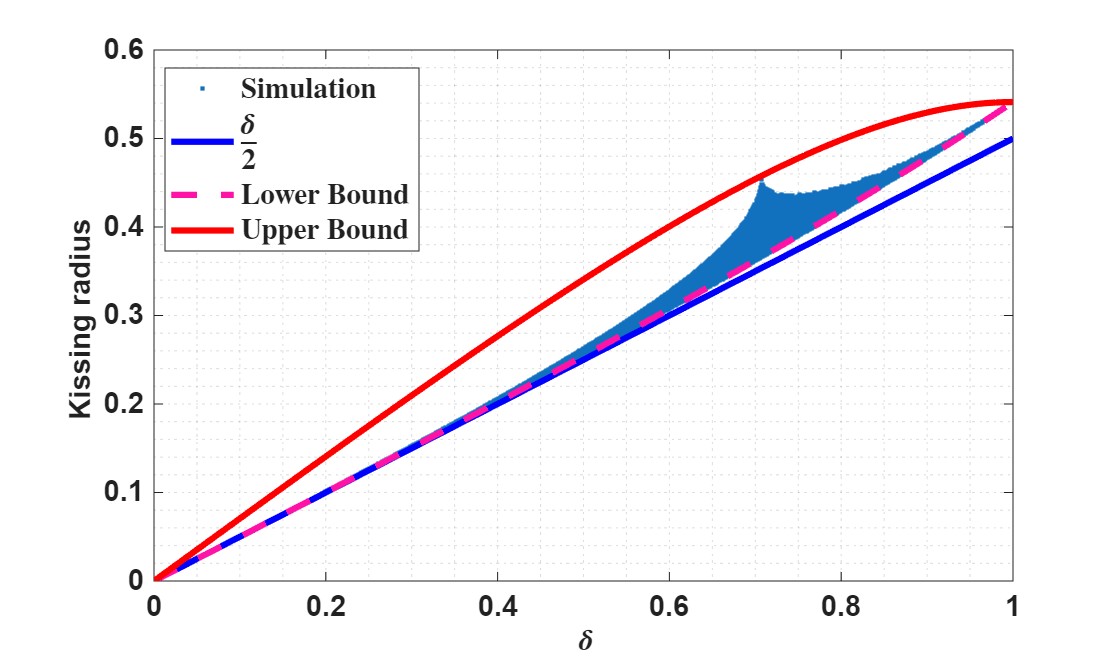}
     \caption{The upper bound $\bar{\varrho}$ and lower bound   $\underline{\varrho}$ of kissing radius in $\mathcal{PU}_4$.  These bounds are compared to simulated midpoints between two randomly generated 
codewords}
     \label{fig:Kissingradius}
 \end{figure}

The upper and lower bounds on the kissing radius in $\PUN$ provide geometric insight into the local packing density of codebooks. By comparing these bounds, we can analyze how tightly the unitary space is covered without overlap, thereby assessing the efficiency and optimality of the constructed projective codebooks.
Fig. \ref{fig:Kissingradius}, for $n=4$  and $500000$ unitary matrices, illustrates the kissing radius bounds provided in Theorem~\ref{Theo:Boundonthekissingradius}. The bounds are compared to simulated midpoints between two randomly generated codewords. It is also compared with the estimate $\frac{\delta}{2}$, corresponding to the classical packing radius in flat geometry. 

\begin{figure}
     \centering
     \includegraphics[width=\linewidth]{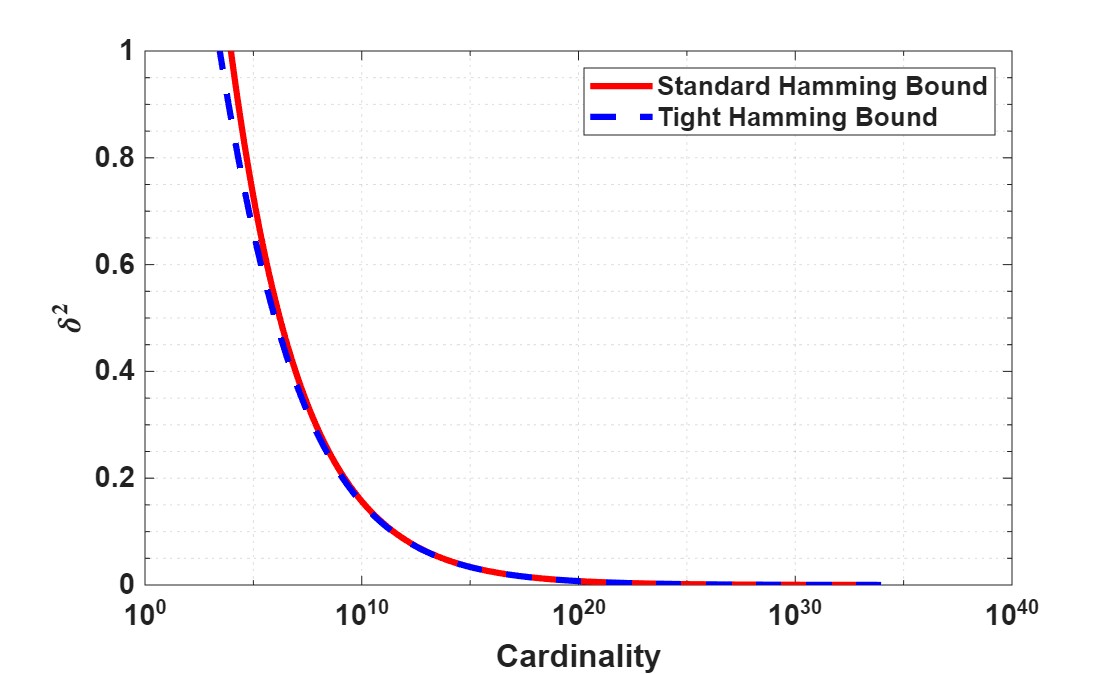}
     \caption{Hamming bound~\eqref{Eq:HammBound} compared with tight Hamming bound ~\eqref{Cor:Lowerboundofkissingradius}  in $\mathcal{PU}_4$.}
     \label{fig:StanderHammingBound}
\end{figure}

In Fig. \ref{fig:StanderHammingBound}, for $n=4$ we  compare the Hamming bound and the tight Hamming bound by kissing radius in terms of rate of codebook and square of minimum distance.  We find that the kissing radius analysis is relevant only for small codebooks. 
 
\begin{figure}
    \centering
    \begin{subfigure}{1\linewidth}
        \centering
        \includegraphics[width=1\linewidth]{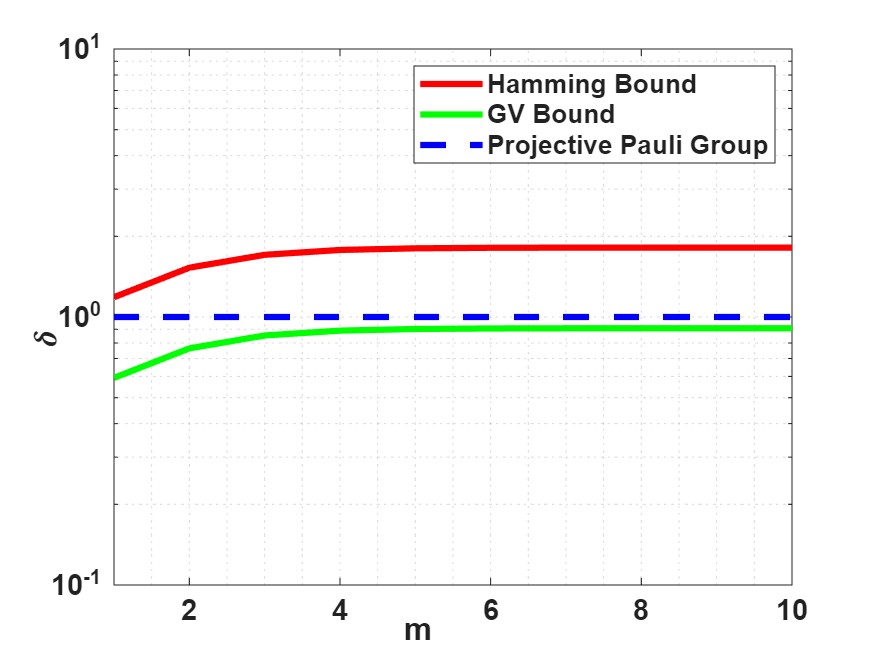}
        \caption{Minimum distance of $\widetilde{\mathcal{P}}_n$, and GV \eqref{Eq:GVBound} and Hamming \eqref{Eq:HammBound} bounds in $n=2^m$ dimensions.}
        \label{fig:ProjectivePauli}
    \end{subfigure}
    
    \begin{subfigure}{1\linewidth}
        \centering
        \includegraphics[width=\linewidth]{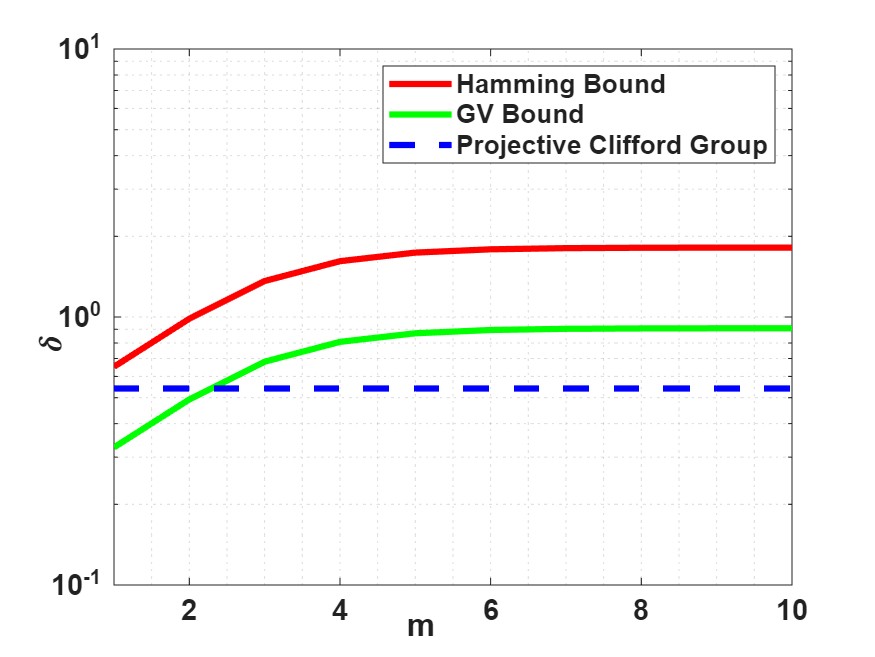}
        \caption{Minimum distance of $\tilde{\mathcal{G}}_n$, and  GV \eqref{Eq:GVBound} and Hamming \eqref{Eq:HammBound} bounds in $n=2^m$ dimensions.}
        \label{fig:projectiveClifford}
    \end{subfigure}
    
    \begin{subfigure}{1\linewidth}
        \centering
        \includegraphics[width=\linewidth]{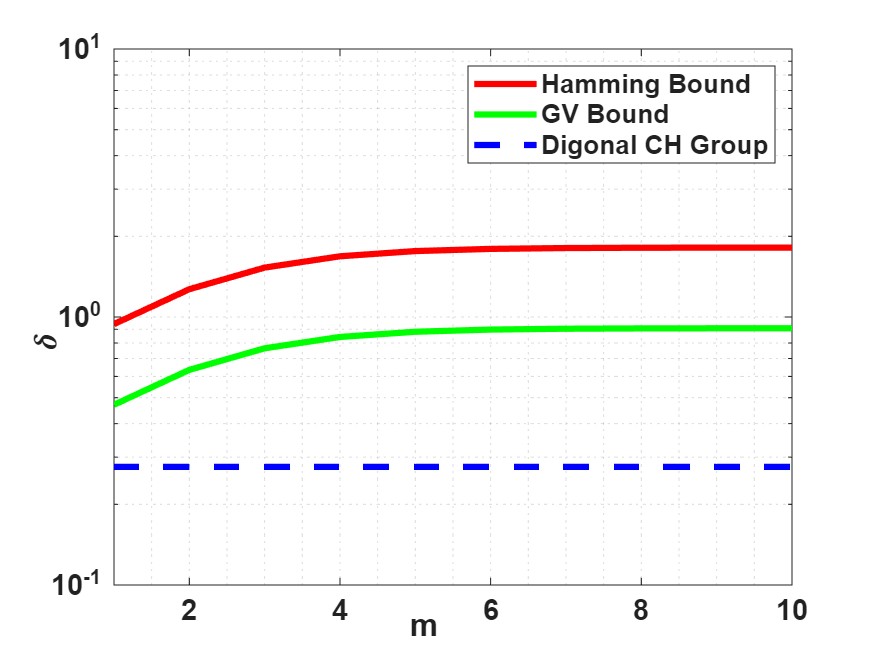}
        \caption{Minimum distance of $\tilde{\mathcal{D}}_{n,3}$, and GV \eqref{Eq:GVBound} and Hamming \eqref{Eq:HammBound}   bounds in $n=2^m$ dimensions.}
        \label{fig:ProjectivDigonalClifordHirarchy}
    \end{subfigure}
    
    \caption{Comparison of minimum distances and theoretical bounds for different code families.}
    \label{fig:all_three}
\end{figure}
In Fig. \ref{fig:all_three}, we compare minimum distances of  codebooks in $\PUN$ with the Hamming \eqref{Eq:HammBound} and  GV \eqref{Eq:GVBound} bounds for the corresponding cardinality. The Hamming bound  of 
provides a strict upper on minimum distance.
The GV bound, in turn shows  that there exists at least one codebook in $\PUN$ whose minimum distance is larger than the   GV. As shown Fig. \ref{fig:ProjectivePauli}, the minimum distance of Pauli matrices  $\widetilde{\mathcal{P}}_n$ lie between these two bounds. The Pauli matrices are optimal due to their otrhoplectic structure. They outperform the general guarantee of the GV bound. In comparison, the Clifford groups of Fig.~\ref{fig:projectiveClifford}, and the diagonal Clifford hierarcy of Fig. \ref{fig:ProjectivDigonalClifordHirarchy} are farther from optimum. For $m=1$ and $m=2$, the Clifford groups outperform the GV-bound. The diagonal Clifford hierarchy is systematically worse than the bound. Packing all the codewords in diagonal matrices compromizes the minimum distance.

\begin{figure}
     \centering
     \includegraphics[width=99mm]{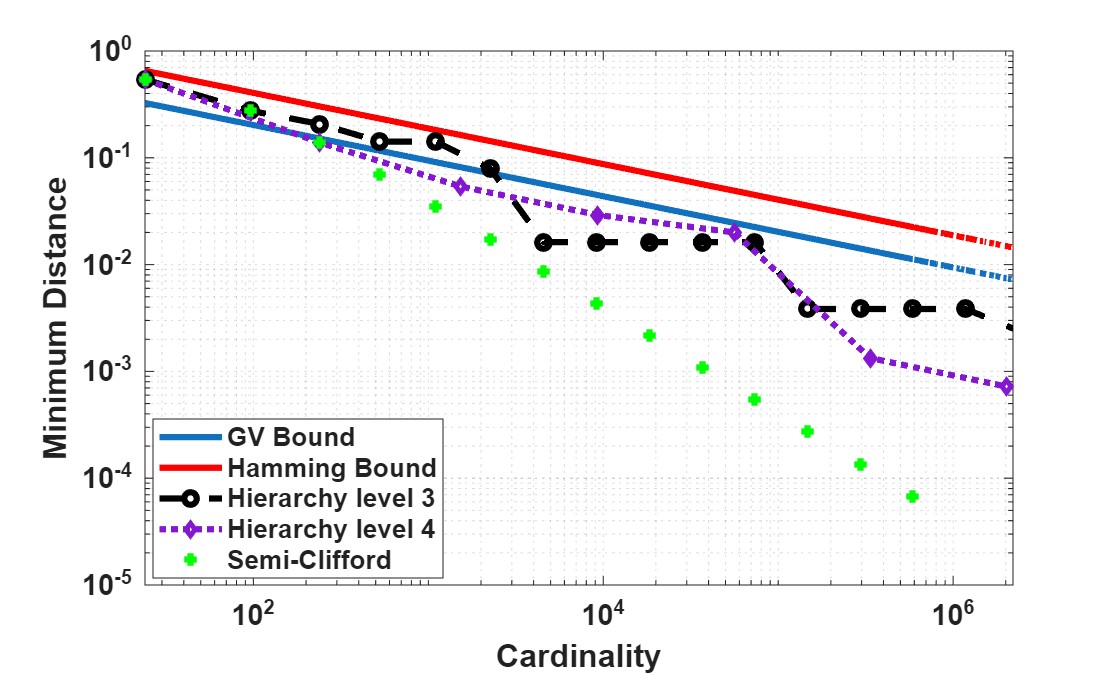}
     \caption{Minimum distances of $\tilde{\mathcal{T}}_l $ , $\tilde{\mathcal{S}}_l $ and $\tilde{\mathcal{C}}_{2,k}$ 
     and comparison with GV \eqref{Eq:GVBound} and Hamming \eqref{Eq:HammBound}  bounds in $\mathcal{PU}_2$.
     }
     \label{fig:Boundcompare1}
\end{figure}

In Fig.~\ref{fig:Boundcompare1}, minimum distances of codebooks of products of higher order Clifford hierarchy elements are shown.  $\tilde{\mathcal{T}}_l $ codes for $l = 0~\text{to}~15$ stages of $\mathbf{T}$-gates and $\tilde{\mathcal{S}}_l $ codes for $l = 0~\text{to}~6$ stages of $\mathbf{S}$-gates in $\mathcal{PU}_2$ are considered. The results are compared to the GV and Hamming bounds. 
The minimum distances are obtained numerically by generating the codebooks and calculated their minimum distances. For small $l$, $\tilde{\mathcal{T}}_l $ outperform the GV bound, while   $\tilde{\mathcal{S}}_l $ is slightly worse than this bound. For larger $l$, the minimum distances generically follow almost a similar slope as the bounds, with certain values of $l$ being considerably better than some other.  

\begin{figure}
     \centering
     \includegraphics[width=99mm]{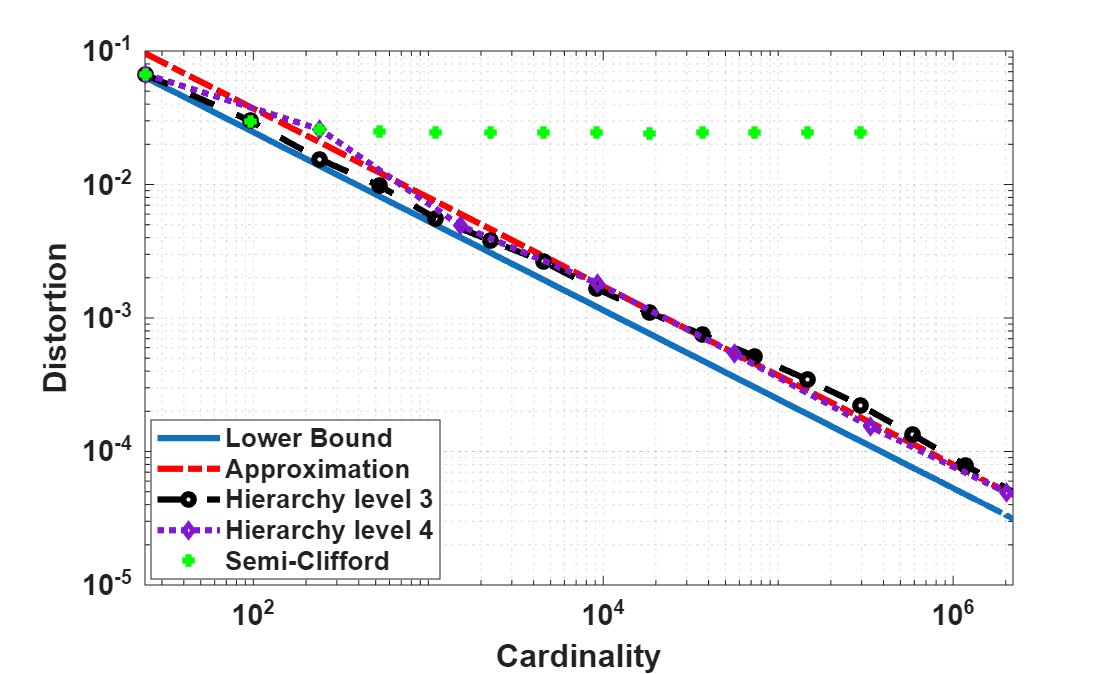}
     \caption{Comparison of distortion of codebooks $\tilde{\mathcal{T}}_l $, $\tilde{\mathcal{S}}_l $, and  codebooks $\tilde{\mathcal{C}}_{2,k}$ 
     with the corresponding bounds \eqref{Eq:GVBound} and \eqref{Eq:HammBound} in $\mathcal{PU}_2$.}
     \label{fig:CTC_CSC_DistorBoundsComp}
\end{figure}

Fig.~\ref{fig:CTC_CSC_DistorBoundsComp} compares the distortion of the $\tilde{\mathcal{T}}_l $, $\tilde{\mathcal{S}}_l $, and semi-Clifford codebooks with the lower and upper bounds of the minimum distortion \eqref{Eq:Distortionbound}
in $\mathcal{PU}_2$. For the semi-Clifford codebook, we considered diagonal parts from levels $k=2, ..., 7$. Here, we considered the quantization distortion of $500000$ random unitary matrices for getting the results. As the figure illustrates, the semi-Clifford codebook exhibits a flooring behavior. After $k=4$, the average distortion is only slightly improved when $k$ increases. 
The reason is in the structure of the semi-Clifford hierarchy---with increasing $k$, there is a increasingly fine codebook of diagonal matrices. The off-diagonal directions, however, are simply given by the Clifford group, they are not getting richer with increasing $l$.  In contrast the codebooks $\tilde{\mathcal{T}}_l $ and $\tilde{\mathcal{S}}_l $ consisting of products of higher level clifford hierarchy elements, exhibit performance very close to the bounds. The upper bound on the optimal distortion is given by the average distortion of random codebooks. Thus we observe that for small $l$, the  $\tilde{\mathcal{T}}_l $ and $\tilde{\mathcal{S}}_l $ are better than a typical random codebook, and for large $l$, they are as good as a typical random codebook.

As the distortion performance of $\tilde{\mathcal{T}}_l $ and $\tilde{\mathcal{S}}_l $ is roughly equal, given the cardinality, the choice of using one rather than the other should be governed by implementation issues. 



\begin{figure}
     \centering
     \includegraphics[width=99mm]{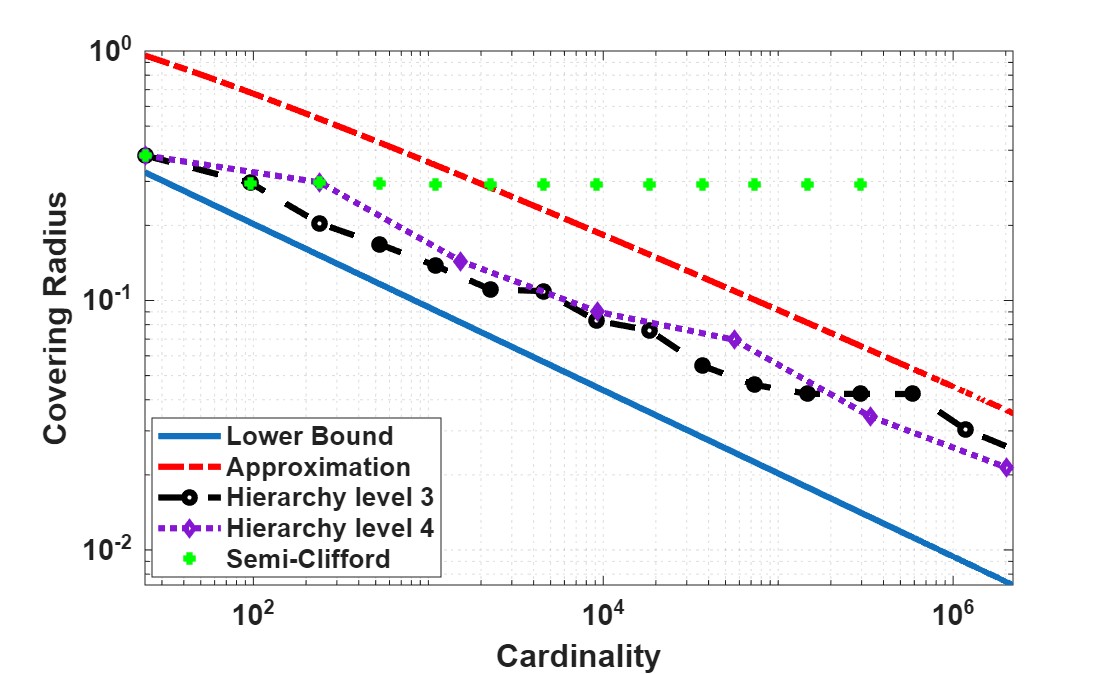}
     \caption{Comparison of the covering radius of codebooks $\tilde{\mathcal{T}}_l $ and $\tilde{\mathcal{S}}_l $, along with the lower bound \eqref{Eq:Lowerboundofcoveringradius} 
     and the approximated covering radius \eqref{Eq:Approximatedcoveringradius} in $\mathcal{PU}_2$.}
     \label{fig:Coveringbounds1}
\end{figure}

The covering radius quantifies the worst-case approximation error between any unitary in $\mathcal{PU}_2$ and its nearest codebook element, thus indicating how uniformly the codebook covers the unitary group space.
In Fig.~\ref{fig:Coveringbounds1} we compare the covering radius of codebooks $\tilde{\mathcal{T}}_l $ for $l=0 ~\text{to}~ 15$ stages of $\mathbf{T}$-gates and $\tilde{\mathcal{S}}_l $ for $l=0~ \text{to}~ 6$ stages of $\mathbf{S}$-gates  with the theoretical lower bound~ \eqref{Eq:Lowerboundofcoveringradius} and the approximated covering radius~ \eqref{Eq:Approximatedcoveringradius}.
To find the  covering radius  of the of these codebooks, we generate 500000 unitary matrices then  find the covering radius using \eqref{Eq:Coviringradius}.
A smaller value of $\rho(\mathcal{C})$ indicates denser coverage and a more uniform sampling of $\mathcal{PU}_2$. 
For small $l$, the covering radius is close to the lower bound. It is interesting to note that for all considered values of $l$, the covering radius of these systematic codebooks are better than the approximate value~\eqref{Eq:Approximatedcoveringradius}, which is valid for random codebooks. 

 \section{Conclusion}\label{sec:8}

In this paper, we considered quantum computation as a coding theoretical problem on the space $\PUN$ of $n\times n$-dimensional projective unitary matrices.
We first calculated the volume of $\PUN$. Using this volume,
we found the measure of small balls in $\PUN$ with respect to the
chordal distance, and established the GV lower
and Hamming upper bounds for codebooks in $\PUN$. In addition, we
provided the upper and lower bounds for the kissing radius of codes in
$\PUN$, which quantifies the maximum radius of non-overlapping metric
balls. Based on normalized volumes of metric balls around the kissing
radius, we established bounds on the density of codes in $\PUN$.
Using the bound on code density, we provided an improved Hamming
bound. Furthermore, we derived lower and upper bounds of the rate--distortion
function over $\PUN$, and provided a lower bound and an approximation for the covering radius. 
As examples of codebooks in $\PUN$, relevant for quantum computation, We considered the projective Pauli group, the projective Clifford group, and the projective diagonal part of the
Clifford hierarchy group, and found
their minimum distances. As examples of larger cardinality codebooks, we considered single-qubit computational codebooks. In addition to higher Clifford-hierarchy level semi-Clifford circuits, we considered codebooks consisting of products of a finite number of 3rd level and 4th level Clifford hierarchy elements. The comparison of numerical performance results to bounds show that increasing the hierarchy level alone does not improve performance much. In contrast, products of multiple higher hierarchy level elements leads to performance which is comparable, and even slightly better than that of random codebooks. 

\section*{Acknowledgment}

 This work was funded in part by Business Finland (grant 8264/31/2022).

\bibliographystyle{IEEEtran}
\bibliography{refs}

\balance

\end{document}